\documentclass[11pt,a4paper]{article}
\usepackage{jcappub}

\usepackage{amsmath,amssymb,amsfonts}
\usepackage{mathtools}
\usepackage{braket}
\usepackage{subfig}

\newcommand{\sd}[1]{\mathrm{#1}}

\title{Boltzmann equations for preheating}

\author[]{William.~T.~Emond,}

\author[]{Peter Millington and}

\author[]{Paul~M.~Saffin}

\affiliation[]{School of Physics and Astronomy, University of Nottingham,\\
University Park, Nottingham NG7 2RD, United Kingdom}

\emailAdd{william.emond@nottingham.ac.uk}
\emailAdd{p.millington@nottingham.ac.uk}
\emailAdd{paul.saffin@nottingham.ac.uk}

\abstract{
We derive quantum Boltzmann equations for preheating by means of the density matrix formalism, which account for both the non-adiabatic particle production and the leading collisional processes between the produced particles. In so doing, we illustrate the pivotal role played by pair correlations in mediating the particle production. In addition, by numerically solving the relevant system of Boltzmann equations, we demonstrate that collisional processes lead to a suppression of the growth of the number density even at the very early stages of preheating.
}


\begin{document}

\maketitle


\section{Introduction}

It has long been known that the Hot Big-Bang (HBB) model cannot provide a complete description of the evolution of the early universe. For instance, it is unable to provide explanations for the horizon problem, the flatness problem and the absence of relic exotic particles (e.g., magnetic monopoles). These observations motivate the need for an additional theory that provides a more complete description. To date, the prevailing theory offering solutions (at least in part) to these problems is that of inflation, initially proposed by Guth in 1980~\cite{Guth:1980zm} and further developed by Linde~\cite{Linde:1981mu}, and Albrecht and Steinhardt~\cite{Albrecht:1982wi}. However, a consequence of the rapid expansion of the early universe driven by the inflaton field $\phi$ is that, once inflation terminated, the universe was super-cooled, with effectively all of the available energy contained in the inflaton potential $V(\phi)$. In the post-inflationary epoch, one is therefore left with the problem of how to transfer this vast potential energy back into the universe in order to enter the HBB phase. This important stage in the evolution of the universe is called reheating~\cite{Abbott:1982hn,Dolgov:1982th,Albrecht:1982mp} (for a review, see Ref.~\cite{Allahverdi:2010xz}).

In early studies of reheating, it was assumed that one could treat the decay of the inflaton field perturbatively, i.e.~that individual quanta of the inflaton field (inflatons) decayed independently of one another to populate the universe with relativistic particles. It was soon realized, however, that this approach ignores non-perturbative effects and, most importantly, the fact that the inflaton field is not a superposition of asymptotically free single inflaton states but rather a coherently oscillating condensate. This coherent behaviour leads to a parametric resonance of the amplitudes of the field modes coupled to the inflaton. As a result, the energy transfer from the inflaton to these fields is extremely rapid and highly non-adiabatic, giving rise to explosive particle production. This phase is referred to as preheating and was first analyzed in detail by Kofman, Linde and Starobinsky~\cite{Kofman:1994rk,Kofman:1997yn} (see also Ref.~\cite{Kofman:1997ga}) and studied further by many others (see, e.g., Refs.~\cite{Dolgov:1989us, Traschen:1990sw, Shtanov:1994ce, Boyanovsky:1995ud, Yoshimura:1995gc, Kaiser:1995fb, Boyanovsky:1996sq}). For a comprehensive review of non-perturbative post-inflationary dynamics, we refer the reader to Ref.~\cite{Amin:2014eta}.

In the case of chaotic inflation~\cite{Linde:1983gd}, and soon after inflation ends, the inflaton condensate $\phi(t)$ undergoes coherent oscillations about the minimum of its potential with an amplitude of order the (reduced) Planck mass. It is during this phase that the inflaton condensate begins to decay through its interactions with other fields, resulting in the production of relativistic particles (see, e.g., Ref.~\cite{Linde:2007fr,Amin:2014eta}). If this production process is sufficiently slow, and the couplings between the fields are sufficiently small, one can treat the dynamics perturbatively~\cite{Podolsky:2005bw}. Moreover, such gradual reheating allows for almost immediate thermalization of the produced particles~\cite{Allahverdi:2010xz}. On the other hand, in the early stages of reheating, non-perturbative processes can dominate, leading to explosive particle production that occurs on time-scales much shorter than those needed for the produced particles to thermalize. Within the chaotic inflation scenario, preheating begins in the broad resonance regime, wherein the particle number grows exponentially across wide bands of momentum~\cite{Kofman:1997yn}. Due to the decay of the amplitude of the inflaton condensate, and the expansion of the universe, the dynamics eventually transition to the narrow resonance regime, where the growth of particle number is restricted to ever narrower bands of momentum.  At this stage, the backreaction from the created particles and the Hubble expansion continue to conspire to reduce the efficiency of the parametric resonance~\cite{Bassett:2005xm}. Specifically, the backreaction alters the structure of the resonance bands, the cosmological expansion redshifts the momenta of the produced particles and both effects cause the created particles to be shifted out of the resonance bands. The resonance can also be blocked by the onset of effective thermal masses~\cite{Lerner:2015uca}. In any case, as the parametric resonance becomes increasingly narrow and inefficient, the dynamics inevitably transition to the perturbative regime~\cite{Kofman:1994rk,Kofman:1997yn}, and it is during this final stage of reheating that scatterings redistribute the occupancy of the momentum modes, leading to the eventual kinetic equilibriation of the primordial plasma. While the time-scales for preheating are much shorter than those needed for thermalization, the processes driving that thermalization are still relevant during the preheating phase. In particular, semi-classical lattice simulations have shown that sufficiently large self-interactions of the produced particles can suppress or prevent the resonant particle production~\cite{Prokopec:1996rr}.

The early thermal history of the universe depends strongly on how the primordial plasma attained kinetic equilibrium~\cite{Podolsky:2005bw}. This has motivated the extensive study of thermalization both in perturbative reheating~\cite{Semikoz:1996ty,Davidson:2000er,Allahverdi:2000ss,Allahverdi:2002pu,Graham:2008vu,Harigaya:2013vwa,Mukaida:2015ria} and after the phase of preheating~\cite{Boyanovsky:1995ema,Son:1996uv,Khlebnikov:1996mc,Kofman:2000rk,Micha:2002ey,Micha:2003ws,Micha:2004bv,Podolsky:2005bw,Kaya:2009yr}, and the relevant relaxation rates can be calculated by means of thermal quantum field theory~\cite{Graham:2008vu,Drewes:2013iaa,Drewes:2014pfa}. The value of the reheat temperature, for instance, has important consequences for leptogenesis (see, e.g., Ref.~\cite{Davidson:2008bu}) and for the generation of dark matter relic densities (see, e.g., Ref.~\cite{Baer:2014eja}). Importantly, the reheat temperature should be larger than a few MeV to allow for the standard Big-Bang Nucleosynthesis (BBN)~\cite{Olive:1999ij}, and requiring that thermalization occur before nucleosynthesis imposes an upper bound on the inflaton mass as a function of the reheat temperature~\cite{McDonald:1999hd}. In the case of supersymmetric theories, the potential over-production of gravitinos~\cite{Khlopov:1984pf,Ellis:1984eq} can spoil the generation of the light elements, providing an upper bound on the reheat temperature. 

The aims of this article are to recast the problem of preheating in the density matrix formalism~\cite{Sigl:1992fn} and to study the impact of scatterings on preheating by means of a system of quantum Boltzmann equations. These Boltzmann equations are able to go beyond the usual mode-function analysis of preheating, based on the Mathieu equation, by accounting simultaneously for both the resonant particle production and the collisional processes. We show that the resonant particle production proceeds via the population of pair correlations of the form \smash{$M_{\mathbf{k}}\sim\langle\hat{a}_{-\mathbf{k}}(t)\hat{a}_{\mathbf{k}}(t)\rangle$} and \smash{$M^*_{\mathbf{k}}\sim\langle\hat{a}^{\dagger}_{\mathbf{k}}(t)\hat{a}^{\dagger}_{-\mathbf{k}}(t)\rangle$}, requiring us to solve the coupled system of Boltzmann equations for the number density  \smash{$N_{\mathbf{k}}\sim\langle\hat{a}_{\mathbf{k}}^{\dag}(t)\hat{a}_{\mathbf{k}}(t)\rangle$} and these ``particle-anti-particle'' correlations. The presence of such pair correlations is expected in the absence of time translational invariance, as was identified in the context of preheating in Refs.~\cite{Herranen:2008hi,Herranen:2008hu,Herranen:2008di} by means of the Schwinger-Keldysh closed-time-path~\cite{Schwinger:1960qe, Keldysh:1964ud} and Kadanoff-Baym~\cite{Baym:1961zz} formalisms of non-equilibrium quantum field theory (see also Refs.~\cite{Blaizot:2001nr,Berges:2004yj}). Therein, the number density must be carefully defined~\cite{Garbrecht:2002pd}, and the pair correlations can then be accounted for in the non-equilibrium Green's functions by working with coherent quasi-particle approximations~\cite{Herranen:2010mh, Herranen:2011zg, Fidler:2011yq} or directly in terms of the operator algebra by means of the so-called interaction-picture approach~\cite{Millington:2012pf,Dev:2014laa}. Particle-anti-particle pair correlations have also been studied in the density matrix formalism in the context of neutrino kinetics, where they may play a role in core-collapse supernovae~\cite{Volpe:2013jgr, Vaananen:2013qja,  Vlasenko:2013fja, Serreau:2014cfa,Kartavtsev:2015eva}.  Here, we show that the pair correlations in fact play the pivotal role in mediating the particle production, and without them no particle production occurs. This leads to a powerful generalization of the previous observation by Morikawa and Sasaki~\cite{Morikawa:1984dz} that small perturbations to such a system would destroy coherences between particle and anti-particle states. Indeed, it follows that any processes that cause such pair correlations to decohere will suppress, or shut off, the resonant particle production.

In order to make the problem tractable and to deal with the time-dependent phase space, we apply a Wigner-Weisskopf approximation~\cite{Weisskopf:1930au}, such that the collisions are effectively treated as occurring only during the periods of adiabatic evolution. Even with this approximation, we find that the lowest-order scatterings have a non-negligible effect on the resonant particle production from the earliest stages of preheating, in agreement with the analysis of Ref.~\cite{Prokopec:1996rr}. In addition, we show that the pair correlations, although of comparable magnitude to the number density, can be neglected in the collision integrals when the collisional processes occur more slowly than the free-phase oscillations of the pair correlations.

The remainder of this article is organized as follows. In Sec.~\ref{sec:preheating}, we provide an overview of preheating and the standard analysis of the resonant particle production by means of the Mathieu equation. We then proceed in Sec.~\ref{sec:densmat} to recast the problem of preheating in the density matrix formalism, deriving quantum Boltzmann equations for the number density and pair correlations. The numerical solution of the resulting system of equations is presented in Sec.~\ref{sec:numerics} for an illustrative set of benchmark parameters. Our conclusions are presented in Sec.~\ref{sec:conc}, and further technical details are included in the appendices.


\section{Parametric resonances and preheating}
\label{sec:preheating}

In order to make a concrete comparison with the density matrix formulation, we shall first provide a brief overview of the standard theory of preheating. We consider a simple toy model described by the Lagrangian density
\begin{equation}\label{lagrangian density}
 \mathcal{L}\ =\ -\:\frac{1}{2}\,\partial^{\mu}\phi\,\partial_{\mu}\phi\:-\:\frac{1}{2}\,m_{\phi}^{2}\,\phi^{2}\:-\:\frac{1}{2}\,\partial^{\mu}\chi\,\partial_{\mu}\chi\:-\:\frac{1}{2}\,m_{\chi}^{2}\chi^{2}\:-\:\frac{g}{4}\,\phi^{2}\chi^{2}\;,
\end{equation}
where $\phi$ is the inflaton field and $\chi$ is a real scalar field, which we might imagine as a proxy for, or being coupled to, the Standard Model fields. While fermionic particle production can occur via parametric resonance (see, e.g., Refs.~\cite{Greene:1998nh,Greene:2000ew,Berges:2010zv}), we restrict our attention to scalar fields to avoid the additional technical complications of dealing with spinor fields. Throughout this article, we work with the metric signature $(-,+,+,+)$.

We proceed by making a mean-field approximation, describing the inflaton condensate by the time-dependent background field $\varphi(t)\coloneqq\braket{\hat{\phi}(t,\mathbf{x})}$ (and assuming $\braket{\hat{\chi}(t,\mathbf{x})}=0$), where $t$ is the cosmic time. The resulting effective Lagrangian for the fluctuations in the $\chi$ field
\begin{flalign}\label{eff lagrangian}
 \mathcal{L}_{\text{eff}}(t,\mathbf{x})\ &=\ 
-\:\frac{1}{2}\,\partial^{\mu}\chi(t,\mathbf{x})\partial_{\mu}\chi(t,\mathbf{x})\:-\:\frac{1}{2}\,m^{2}_{\text{eff}}(t)\chi^{2}(t,\mathbf{x})
\end{flalign}
then contains a time-dependent mass
\begin{equation}\label{eff mass}
m^{2}_{\text{eff}}(t)\ =\ m_{\chi}^{2}\:+\:g\,\varphi^{2}(t)/2\;.
\end{equation}
Note that we have omitted the interactions between the $\chi$ and inflaton fluctuations, which give rise to perturbative decays that play a subdominant role in the particle production. The modes of the $\chi$ fluctuations evolve according to
\begin{equation}\label{standard preheat}
 \ddot{\chi}_{\mathbf{k}}(t)\:+\:3H(t)\dot{\chi}_{\mathbf{k}}(t)\:+\:\omega_{\mathbf{k}}^{2}(t)\chi_{\mathbf{k}}(t)\ =\ 0\;,
\end{equation}
where $H(t)\coloneqq\dot{a}(t)/a(t)$ is the Hubble rate, $a(t)$ is the scale factor and $\omega_{\mathbf{k}}^{2}(t) = \mathbf{k}^{2}/a^2(t)+m_{\text{eff}}^2(t)$.

The evolution of the background value of the inflaton field $\varphi(t)$ is described by the following equations of motion:
\begin{subequations}\label{inflaton eoms}
 \begin{gather}
\label{inflaton eom 1} 
\ddot{\varphi}(t)\:+\:3H(t)\dot{\varphi}(t)\:+\:m^{2}_{\phi}\,\varphi(t)\ =\ 0\;,\\
\label{inflaton eom 2}
3M_{\rm Pl}^2H^{2}(t)\ =\ \frac{1}{2}\,\dot{\varphi}^{2}\:+\:V(\varphi)\;,
\end{gather}
\end{subequations}
where $M_{\text{Pl}}$ is the reduced Planck mass. At the start of reheating, the solution to Eqs.~\eqref{inflaton eom 1} and \eqref{inflaton eom 2} asymptotically approaches~\cite{Kofman:1994rk}
\begin{equation}
\label{inflaton sol}
\varphi(t)\ =\ \varphi_{0}(t)\cos(m_{\phi}t)\;,
\end{equation}
where $\varphi_{0}(t) =2\sqrt{6}M_{\text{Pl}}/(3m_{\phi}t)$ is a slowly decaying amplitude. This solution neglects the dissipative effects of particle production and its backreaction on the inflaton condensate, which increases the rate of decay. However, this backreaction is expected to be subdominant in the early stages of preheating that we study here~\cite{Kofman:1994rk,Kofman:1997yn}.

Assuming that the rate of post-inflationary expansion is small relative to the rates of particle production and thermalization, we can take $\dot{a}\approx0$ and $\varphi_0(t)\approx\text{const}$. We can then recast the mode equation~\eqref{standard preheat} in the form of a Mathieu equation as
\begin{equation}
\label{mathieu}
\chi''_{\mathbf{k}}(z)\:+\:\big[A_{\mathbf{k}}\:-\:2q\cos(2z)\big]\chi_{\mathbf{k}}(z)\ =\ 0\;,
\end{equation}
where $z \coloneqq m_{\phi}t+\pi/2$, $A_{k}\coloneqq(\mathbf{k}^{2}+m_{\chi}^{2})/m_{\phi}^{2}+2q$ and $q\coloneqq g\varphi_{0}^{2}/(8m_{\phi}^{2})$.  An important feature of the solutions to Eq.~\eqref{mathieu} is that, within certain regions of momentum space,\footnote{Note that for narrow resonance, i.e.~$q\ll1$, these regions of instability occur within certain narrow bands of frequencies $\Delta k^{(n)}$ (labelled by an integer index $n$), corresponding to a set of instability parameters $\mu_{\mathbf{k}}^{(n)}$. In the case of broad resonance, i.e.~$q\gg1$, the instability occurs for a wide range of momenta, across a continuous spectrum, and the amplification is much more efficient than in the case of narrow resonance.} there exist exponential instabilities $\chi_{\mathbf{k}}(z)\,\propto\,\exp(\mu_{\mathbf{k}}z)$, whose growth rate is parameterized by the instability parameter $\mu_{\mathbf{k}}$. The latter depends on both $q$ and $A_{\mathbf{k}}$ and is given by
\begin{equation}
\mu_{\mathbf{k}}\ =\ \frac{1}{\pi}\,\ln\bigg|\sqrt{F_{\mathbf{k}}^{2}}+\sqrt{F_{\mathbf{k}}^{2}-1}\bigg|\;,
\end{equation}
where
\begin{equation}
F_{\mathbf{k}}\ =\ 1\:+\:\bigg[\frac{\sd{d}}{\sd{d}z}\,\tilde{\chi}^{(1)}_{\mathbf{k}}(z=\pi/2)\bigg]\tilde{\chi}^{(2)}_{\mathbf{k}}(z=\pi/2)\;,
\end{equation}
and $\tilde{\chi}^{(1)}_{\mathbf{k}}$ and $\tilde{\chi}^{(2)}_{\mathbf{k}}$ are solutions to Eq.~\eqref{mathieu}, satisfying the initial conditions $\{\tilde{\chi}^{(1)}_{\mathbf{k}}=1$, $\sd{d}\tilde{\chi}^{(1)}_{\mathbf{k}}/\sd{d}z=0\}$ and $\{\tilde{\chi}^{(2)}_{\mathbf{k}}=0$, $\sd{d}\tilde{\chi}^{(2)}_{\mathbf{k}}/\sd{d}z=1\}$ at $z=0$~\cite{Taruya:1998cz,abramowitz1964handbook}. Note that parametric resonance occurs whenever $F_{\mathbf{k}}>1$.

These instabilities lead to the exponential growth of occupation numbers of quantum fluctuations \smash{$N_{\mathbf{k}}(t)\propto\exp(2\mu_{\mathbf{k}}^{(n)}z)$}. This particle production is a consequence of the effective frequency $\omega_{\mathbf{k}}(t)$ repeatedly violating the adiabaticity condition~\cite{Kofman:1997yn}
\begin{equation}
\label{eq:adiabaticapprox}
\bigg|\frac{\dot{\omega}_{\mathbf{k}}(t)}{\omega^{2}_{\mathbf{k}}(t)}\bigg|\ \ll\ 1
\end{equation}
as $\varphi(t)$ passes through the minimum of its potential at $\varphi(t)=0$.

The occupation number can be extracted in one of two ways:
\begin{itemize}

\item [(i)] The effective Hamiltonian can be written in the form
\begin{equation}
\hat{H}_{\text{eff}}(t)\ =\ \int\!\frac{\sd{d}^{3}\mathbf{k}}{(2\pi)^{3}}\;\hat{H}_{\mathbf{k}}^{0}(t)\ =\ {\rm Vol}\int\!\frac{\sd{d}^{3}\mathbf{k}}{(2\pi)^{3}}\;\omega_{\mathbf{k}}(t)\bigg[\hat{N}_{\mathbf{k}}(t)\:+\:\frac{1}{2}\bigg]\;,
\end{equation}
where 
\begin{equation}\label{n def}
 \hat{N}_{\mathbf{k}}(t)\ =\ \frac{\hat{a}^{\dagger}_{\mathbf{k}}(t)\hat{a}_{\mathbf{k}}(t)}{\text{Vol}}
\end{equation}
is the number operator and $\text{Vol}=(2\pi)^{3}\delta^{(3)}(\mathbf{0})$. The energy per momentum mode is then given as follows:
\begin{equation}\label{hamiltonian num density}
 \langle\hat{H}_{\mathbf{k}}^{0}(t)\rangle\ =\ \frac{1}{2}\,\lvert\dot{\chi}_{\mathbf{k}}\rvert^{2}\:+\:\frac{1}{2}\,\omega^{2}_{\mathbf{k}}(t)\lvert\chi_{\mathbf{k}}\rvert^{2}\ =\ {\rm Vol}\,\omega_{\mathbf{k}}(t)\bigg[N_{\mathbf{k}}(t)\:+\:\frac{1}{2}\bigg]\; ,
\end{equation}
and we can therefore extract the occupation number $N_{\mathbf{k}}(t)=\braket{\hat{N}_{\mathbf{k}}(t)}$ (per momentum mode) as~\cite{Kofman:1997yn}
\begin{equation}
N_{\mathbf{k}}(t)\ =\ \frac{1}{2\,\text{Vol}\,\omega_{\mathbf{k}}(t)}\bigg[\lvert\dot{\chi}_{\mathbf{k}}\rvert^{2}\:+\:\omega^{2}_{\mathbf{k}}(t)\lvert\chi_{\mathbf{k}}\rvert^{2}\bigg]\:-\:\frac{1}{2}\;.
\end{equation}
The total number per unit volume $N(t)$ is then obtained by integrating over all momentum space, i.e.
\begin{equation}
 N(t)\ =\ \int\!\frac{\mathrm{d}^{3}\mathbf{k}}{(2\pi)^{3}}\;N_{\mathbf{k}}(t)\;.
\end{equation}

\item [(ii)] The second way is to consider the Bogoliubov transformation~\cite{Bogolyubov:1958kj}
\begin{subequations}\label{aa* ansatz}
\begin{align}
 \hat{a}_{\mathbf{k}}(t)\ &=\ \alpha_{\mathbf{k}}(t,t')\hat{a}_{\mathbf{k}}(t')\:+\:\beta_{\mathbf{k}}^{\ast}(t,t')\hat{a}^{\dagger}_{-\mathbf{k}}(t')\;,\\ \hat{a}^{\dagger}_{\mathbf{k}}(t)\ &=\ \alpha_{\mathbf{k}}^{\ast}(t,t')\hat{a}^{\dagger}_{\mathbf{k}}(t')\:+\:\beta_{\mathbf{k}}(t,t')\hat{a}_{-\mathbf{k}}(t')\;,
\end{align}
\end{subequations}
where $\alpha_{\mathbf{k}}(t,t')$ and $\beta_{\mathbf{k}}(t,t')$ satisfy
\begin{equation}
 \alpha_{\mathbf{k}}(t,t')\ =\ \alpha_{-\mathbf{k}}(t,t')\;,\qquad\beta_{\mathbf{k}}(t,t')\ =\ \beta_{-\mathbf{k}}(t,t')\;,
\end{equation}
along with the following boundary conditions:
\begin{equation}\label{aa* initial conditions}
 \alpha_{\mathbf{k}}(t,t)\ =\ 1\; ,\quad\beta_{\mathbf{k}}(t,t)\ =\ 0\;.
\end{equation}
Since the Bogoliubov transformation mixes creation and annihilation operators, the vacuum state at the initial time $\lvert 0_{0}\rangle$ appears to the annihilation operator at the later time as an excited state with a non-vanishing occupation number given by
\begin{equation}\label{bogoliubov num density}
 N_{\mathbf{k}}(t)\ =\ \frac{1}{\text{Vol}}\,\langle 0_{0}\rvert\hat{a}^{\dagger}_{\mathbf{k}}(t)\hat{a}_{\mathbf{k}}(t)\rvert 0_{0}\rangle\ =\ \lvert\beta_{\mathbf{k}}(t)\rvert^{2}\;,
\end{equation}
where $\beta_{\mathbf{k}}(t):=\beta_{\mathbf{k}}(t,0)$. 

\end{itemize}

The above descriptions do not take into account the thermalization that may begin to take place during the process of preheating. In order to do so, one must go beyond the mode analysis, and we will describe an approach based on the density matrix formalism in the sections that follow.


\section{Preheating in the density matrix formalism}
\label{sec:densmat}

In this section, we will reformulate the problem of preheating in the so-called density matrix formalism~\cite{Sigl:1992fn}. This will enable us to derive a self-consistent set of quantum Boltzmann equations that describe the evolution of the (scalar) particle number densities throughout preheating. In particular, we will be able to account for collisional processes, which we take to arise from a quartic self-interaction potential for the $\chi$ field
\begin{equation}
\label{int lag}
 \mathcal{L}^{\rm eff}_{\text{int}}(t,\mathbf{x})\ =\ -\:\frac{\lambda}{4!}\,\chi^{4}(t,\mathbf{x})\;,
\end{equation}
which we append to the effective Lagrangian density in Eq.~\eqref{eff lagrangian}, where $\lambda$ is a dimensionless coupling constant. 

We note that the additional effective interaction \smash{$\mathcal{L}^{\rm eff}_{\rm int}\supset -g\varphi(t)\phi\chi^{2}/2$}, generated from the coupling between $\phi$ and $\chi$, gives rise to $\phi$ mediated two-to-two scatterings of $\chi$ particles. An immediate concern is that these processes may dominate over those arising from Eq.~\eqref{int lag}. However, a comparison of the cross-sections of the two different scattering processes reveals that this is not the case for a range of suitable couplings. Let us consider 
the two regimes of each oscillation interval of $\varphi$, namely the intervals of adiabaticity and non-adiabaticity. In the adiabatic regimes, the momenta of the produced particles will typically be $|\mathbf{k}|\lesssim m_{\phi}$, in which case the centre-of-mass energy scales as $\sqrt{s}\sim \sqrt{2g}\varphi_{0}$. One can then show that the $t$- and $u$-channels dominate the $\phi$-mediated processes.\footnote{Here, we consider only tree-level processes, as higher-order contributions will be loop suppressed.} However, their combined cross-section is still suppressed by $\mathcal{O}(10^{-3})$ relative to that of the self-interaction for the values of $g\sim 10^{-7}$ and $\lambda\sim 0.1$ that we take in this analysis (see section~\ref{sec:numerics}). In the non-adiabatic regimes, the time-dependent coupling $g\varphi(t)\to 0$, and the total cross-section vanishes (even for the smallest centre-of-mass energies $\sqrt{s}\to 2m_{\chi}$). We are therefore safe to proceed under the assumption that the dominant collisional processes arise from Eq.~\eqref{int lag}. 

\subsection{Canonical quantization with a time-dependent mass}

We canonically quantize the $\chi$ field in the presence of a time-dependent mass by specifying that $\hat{\chi}(t,\mathbf{x})$ and its canonical conjugate momentum $\hat{\pi}_{\chi}(t,\mathbf{x})$ satisfy the following equal-time commutation relations:
\begin{subequations}
\begin{gather}\label{canon comm}
 \big[\hat{\chi}(t,\mathbf{x}),\hat{\chi}(t,\mathbf{y})\big]\ =\ 0 \ = \ \big[\hat{\pi}_{\chi}(t,\mathbf{x}),\hat{\pi}_{\chi}(t,\mathbf{y})\big]\;,\\\big[\hat{\chi}(t,\mathbf{x}),\hat{\pi}_{\chi}(t,\mathbf{y})\big]\ =\ i\delta^{(3)}\left(\mathbf{x}-\mathbf{y}\right)\;.
\end{gather}
\end{subequations}
This canonical approach was also applied in Ref.~\cite{Morikawa:1984dz}.

We work in the (modified) interaction picture and expand $\hat{\chi}(t,\mathbf{x})$ in terms of its Fourier modes as
\begin{equation} \label{field expansion}
 \hat{\chi}(t,\mathbf{x})\ =\ \int_{\mathbf{k}}\;\Big[\hat{a}_{\mathbf{k}}(t)\,e^{+i\mathbf{k}\cdot\mathbf{x}}\:+\:\hat{a}^{\dagger}_{\mathbf{k}}(t)\,e^{-i\mathbf{k}\cdot\mathbf{x}}\Big]\;,
\end{equation}
where the operators $\hat{a}_{\mathbf{k}}^{\dag}(t)\coloneqq\hat{a}^{\dagger}(t,\mathbf{k})$ and $\hat{a}_{\mathbf{k}}(t)\coloneqq\hat{a}(t,\mathbf{k})$ create and destroy quanta with instantaneous frequency $\omega_{\mathbf{k}}(t)$, given by $\omega^{2}_{\mathbf{k}}(t)=\mathbf{k}^{2}+m_{\text{eff}}^{2}(t)$ (for the scale factor $a\approx \text{const.}=1$). Throughout this article, we use the shorthand notation
\begin{equation}
\label{eq:measure}
\int_{\mathbf{k}}\,\coloneqq\,\int\frac{\mathrm{d}^{3}\mathbf{k}}{(2\pi)^{3}}\;\frac{1}{\sqrt{2\omega_{\mathbf{k}}(t)}}\;.
\end{equation}
The creation and annihilation operators have mass dimension $-\,3/2$ and satisfy the canonical commutation relation
\begin{equation}
\label{eq:CCR}
\big[\hat{a}_{\mathbf{k}}(t),\hat{a}^{\dag}_{\mathbf{k}'}(t)\big]\ =\ (2\pi)^3\delta^3(\mathbf{k}-\mathbf{k}')\;,
\end{equation}
with all other commutators vanishing. This convention has the advantage that the right-hand side of the Eq.~\eqref{eq:CCR} is time-independent.

The creation and annihilation operators $\hat{a}_{\mathbf{k}}(t)$ and $\hat{a}^{\dagger}_{\mathbf{k}}(t)$ can be expressed in terms of the field $\hat{\chi}(t,\mathbf{x})$ and its conjugate momentum $\hat{\pi}_{\chi}(t,\mathbf{x})$ as
\begin{subequations}\label{create/annihilate}
\begin{flalign} 
\hat{a}_{\mathbf{k}}(t)\ &=\ \frac{1}{\sqrt{2}}\int_{\mathbf{x}}\;e^{-i\mathbf{k}\cdot\mathbf{x}}\;\Big[\omega^{1/2}_{\mathbf{k}}(t)\,\hat{\chi}(t,\mathbf{x})\:+\:i\,\omega_{\mathbf{k}}^{-1/2}(t)\,\hat{\pi}_{_{\chi}}(t,\mathbf{x})\Big]\;,\label{annihilate}\\
\hat{a}^{\dagger}_{\mathbf{k}}(t)\ &=\ \frac{1}{\sqrt{2}}\int_{\mathbf{x}}\;e^{+i\mathbf{k}\cdot\mathbf{x}}\;\Big[\omega^{1/2}_{\mathbf{k}}(t)\,\hat{\chi}(t,\mathbf{x})\:-\:i\,\omega^{-1/2}_{\mathbf{k}}(t)\,\hat{\pi}_{_{\chi}}(t,\mathbf{x})\Big]\label{create}\;.
\end{flalign}
\end{subequations}
Since $\hat{\chi}(x)$ and $\hat{\pi}_{\chi}(x)$ are canonical variables, they have no explicit time-dependence. This is not true of the creation and annihilation operators: the time-dependence of the effective (instantaneous) frequency generates an explicit time-dependence for $\hat{a}_{\mathbf{k}}(t)$ and $\hat{a}^{\dagger}_{\mathbf{k}}(t)$. Their full time-evolution is governed by the (interaction-picture) Heisenberg equations
\begin{subequations}\label{time evo aa*}
\begin{flalign}
 \dot{\hat{a}}_{\mathbf{k}}(t)\ =& \ i\left[\hat{H}_{0}(t),\hat{a}_{\mathbf{k}}(t)\right]\:+\:\frac{1}{2}\,\frac{\dot{\omega}_{\mathbf{k}}(t)}{\omega_{\mathbf{k}}(t)}\,\hat{a}^{\dagger}_{-\mathbf{k}}(t)\;,\\ \dot{\hat{a}}^{\dagger}_{\mathbf{k}}(t)\ =&\ i\left[\hat{H}_{0}(t),\hat{a}^{\dagger}_{\mathbf{k}}(t)\right]\:+\:\frac{1}{2}\,\frac{\dot{\omega}_{\mathbf{k}(t)}}{\omega_{\mathbf{k}}(t)}\,\hat{a}_{-\mathbf{k}}(t)\;,
\end{flalign}
\end{subequations}
where the right-most terms have resulted from the time-dependence of the free effective Hamiltonian. We see that the explicit time-dependence mixes the creation and annihilation operators, and this will result in non-trivial pair correlations that are essential to the non-adiabatic particle production of preheating.  

The effective (normal-ordered) Hamiltonian operator $\hat{H}^{\rm eff}_{0}(t)$ can be written as
\begin{equation}\label{quantum H}
 \hat{H}^{\rm eff}_{0}(t)\ =\ \int\!\frac{\mathrm{d}^{3}\mathbf{k}}{(2\pi)^{3}}\;\omega_{\mathbf{k}}(t)\hat{a}^{\dagger}_{\mathbf{k}}(t)\hat{a}_{\mathbf{k}}(t)\;,
\end{equation}
from which it follows that
\begin{subequations}
\label{eq:aadagfull}
\begin{flalign}
 \dot{\hat{a}}_{\mathbf{k}}(t)\ &=\ -\:i\omega_{\mathbf{k}}(t)\hat{a}_{\mathbf{k}}(t)\:+\:\frac{1}{2}\,\frac{\dot{\omega}_{\mathbf{k}}(t)}{\omega_{\mathbf{k}}(t)}\,\hat{a}^{\dagger}_{-\mathbf{k}}(t)\;,\\ \dot{\hat{a}}^{\dagger}_{\mathbf{k}}(t)\ &=\ +\:i\omega_{\mathbf{k}}(t)\hat{a}^{\dagger}_{\mathbf{k}}(t)\:+\:\frac{1}{2}\,\frac{\dot{\omega}_{\mathbf{k}(t)}}{\omega_{\mathbf{k}}(t)}\,\hat{a}_{-\mathbf{k}}(t)\;.
\end{flalign}
\end{subequations}
For completeness, we note that the canonical momentum operator
\begin{equation}\label{canon momentum}
 \hat{\pi}_{\chi}(t,\mathbf{x})\ = \ i\int_{\mathbf{k}}\;\omega_{\mathbf{k}}(t)\left[\hat{a}^{\dagger}_{\mathbf{k}}(t)e^{-i\mathbf{k}\cdot\mathbf{x}}\:-\:\hat{a}_{\mathbf{k}}(t)e^{+i\mathbf{k}\cdot\mathbf{x}}\right]
\end{equation}
is consistent with $\hat{\pi}_{\chi}(x)\,=\,\dot{\hat{\chi}}(x)$. 

Ultimately, we are interested in deriving rate equations that describe the evolution of the number density of $\chi$ quanta. Due to the mixing of the creation and annihilation operators in Eq.~\eqref{eq:aadagfull}, the number operator $\hat{N}_{\mathbf{k}}(t)$, defined in Eq.~\eqref{n def}, has an explicit time-dependence, evolving according to\footnote{$\text{Re}$ gives the Hermitian part for operator-valued arguments.}
\begin{equation}
\label{eq:ndotsource}
 \dot{\hat{N}}_{\mathbf{k}}(t)\ =\ \frac{\dot{\omega}_{\mathbf{k}}(t)}{\omega_{\mathbf{k}}(t)}\,\text{Re}\,\hat{M}_{\mathbf{k}}(t)\;,
\end{equation}
where we have defined the pair operators
\begin{subequations}
\begin{align}
 \hat{M}_{\mathbf{k}}(t)\ &\coloneqq\ \frac{\hat{a}_{\mathbf{-k}}(t)\hat{a}_{\mathbf{k}}(t)}{\text{Vol}}\;,\\ \hat{M}^{\dagger}_{\mathbf{k}}(t)\ &\coloneqq\ \frac{\hat{a}^{\dagger}_{\mathbf{k}}(t)\hat{a}^{\dagger}_{\mathbf{-k}}(t)}{\text{Vol}}\;,
\end{align}
\end{subequations}
which evolve according to
\begin{equation}
\label{M op evo}
 \dot{\hat{M}}_{\mathbf{k}}(t)\ =\ -\:2i\omega_{\mathbf{k}}(t)\hat{M}_{\mathbf{k}}(t)\:+\:\frac{1}{2}\,\frac{\dot{\omega}_{\mathbf{k}}(t)}{\omega_{\mathbf{k}}(t)}\,\big(2\hat{N}_{\mathbf{k}}(t)\:+\:1\big)\;.
\end{equation}
Notice that if there were no non-adiabaticity in the evolution of the system, the expectation values of the pair operators $\hat{M}_{\mathbf{k}}(t)$ and  $\hat{M}^{\dagger}_{\mathbf{k}}(t)$ would vanish. In this case, particle production would cease, and the number density 
\begin{equation}\label{num density exp}
 N_{\mathbf{k}}(t)\ \coloneqq\ \big\langle\hat{N}_{\mathbf{k}}(t)\big\rangle_{t}\ =\ \text{Tr}\,\big[\hat{\rho}(t)\hat{N}_{\mathbf{k}}(t)\big]
\end{equation}
would become an adiabatic invariant. We take the density operator $\hat{\rho}(t)$ to be normalized to unity, i.e.~$\mathrm{Tr}\,\hat{\rho}(t)\ =\ 1$.


\subsection{Master equations}
\label{ssec:master eqs}

The time-evolution of the number density is given by
\begin{equation}
\label{eq:ndotfirst}
 \dot{N}_{\mathbf{k}}(t)\ =\ \frac{\mathrm{d}}{\mathrm{d}t}\,\text{Tr}\,\big[\hat{\rho}(t)\hat{N}_{\mathbf{k}}(t)\big]\ =\ \text{Tr}\,\big[\hat{\rho}(t)\dot{\hat{N}}_{\mathbf{k}}(t)\big]\:+\:\text{Tr}\,\big[\dot{\hat{\rho}}(t)\hat{N}_{\mathbf{k}}(t)\big]\;,
\end{equation}
and the density operator $\hat{\rho}(t)$ evolves according to the quantum Liouville equation
\begin{equation}
\label{density op evo}
 \dot{\hat{\rho}}(t)\ =\ -\:i\big[\hat{H}_{\text{int}}(t),\hat{\rho}(t)\big]\;.
\end{equation}
Integrating Eq.~\eqref{density op evo} with respect to time, applying the method of successive substitution and finally differentiating again with respect to $t$, we can recast the time-evolution of $\hat{\rho}(t)$ in the form
\begin{equation}\label{density op evo trick}
 \dot{\hat{\rho}}(t)\ =\ -\:i\big[\hat{H}_{\text{int}}(t),\hat{\rho}(t_{0})\big]\:-\:\int_{t_{0}}^{t}\mathrm{d}t'\;\big[\hat{H}_{\text{int}}(t),\big[\hat{H}_{\text{int}}(t'),\hat{\rho}(t')\big]\big]\;.
\end{equation}
Here, $t_{0}$ is the intitial time. Substituting this result into Eq.~\eqref{eq:ndotfirst} and making use of Eq.~\eqref{eq:ndotsource}, we have
\begin{flalign}
 \dot{N}_{\mathbf{k}}(t)\ &=\ \frac{\dot{\omega}_{\mathbf{k}}(t)}{\omega_{\mathbf{k}}(t)}\,\text{Re}\,M_{\mathbf{k}}(t)\:-\:i\,\text{Tr}\big\{\big[\hat{H}_{\text{int}}(t),\hat{\rho}(t_{0})\big]\hat{N}_{\mathbf{k}}(t)\big\}\nonumber\\ &\qquad -\:\int_{t_{0}}^{t}\mathrm{d}t'\;\text{Tr}\big\{\big[\hat{H}_{\text{int}}(t),\big[\hat{H}_{\text{int}}(t'),\hat{\rho}(t')\big]\big]\hat{N}_{\mathbf{k}}(t)\big\}\;,
\end{flalign}
where $M_{\mathbf{k}}(t)\coloneqq\text{Tr}\big[\hat{\rho}(t)\hat{M}_{\mathbf{k}}(t)\big]$ and $M^{\ast}_{\mathbf{k}}(t)\coloneqq\text{Tr}\big[\hat{\rho}(t)\hat{M}^{\dagger}_{\mathbf{k}}(t)\big]$.
Using the cyclicity property of the trace, this can be rewritten as
\begin{equation}\label{num density full evo}
 \dot{N}_{\mathbf{k}}(t)\ =\ \frac{\dot{\omega}_{\mathbf{k}}(t)}{\omega_{\mathbf{k}}(t)}\,\text{Re}\,M_{\mathbf{k}}(t)\:-\:i\,\big\langle\big[\hat{N}_{\mathbf{k}}(t),\hat{H}_{\text{int}}(t)\big]\big\rangle_{t_{0}}\:-\:\int_{t_{0}}^{t}\mathrm{d}t'\;\big\langle\big[\big[\hat{N}_{\mathbf{k}}(t),\hat{H}_{\text{int}}(t)\big],\hat{H}_{\text{int}}(t')\big]\big\rangle_{t'}\;.
\end{equation}

Proceeding similarly for $M_{\mathbf{k}}(t)$ and $M^{\dag}_{\mathbf{k}}(t)$, we obtain the following evolution equation for the pair correlation $M_{\mathbf{k}}(t)$
\begin{flalign}\label{M full evo}
 \dot{M}_{\mathbf{k}}(t)\ &=\ -\:2i\omega_{\mathbf{k}}(t)M_{\mathbf{k}}(t)\:+\:\frac{1}{2}\,\frac{\dot{\omega}_{\mathbf{k}}(t)}{\omega_{\mathbf{k}}(t)}\,\big(2N_{\mathbf{k}}(t)\:+\:1\big)\:-\:i\,\big\langle\big[\hat{M}_{\mathbf{k}}(t),\hat{H}_{\text{int}}(t)\big]\big\rangle_{t_{0}}\nonumber \\ & \qquad-\:\int_{t_{0}}^{t}\mathrm{d}t'\;\big\langle\big[\big[\hat{M}_{\mathbf{k}}(t),\hat{H}_{\text{int}}(t)\big],\hat{H}_{\text{int}}(t')\big]\big\rangle_{t'}\;,
\end{flalign}
along with its complex conjugate.

Equations~\eqref{num density full evo} and \eqref{M full evo} compose a coupled set of self-consistent Boltzmann equations that describe the complete evolution of the system, including non-Markovian memory effects. The latter make the solution of this system technically challenging. However, under the assumption that the non-Markovian effects are subdominant, we can make the system tractable by means of a Wigner-Weisskopf (or Markovian) approximation~\cite{Weisskopf:1930au}. This relies on two assumptions:
\begin{enumerate}

\item molecular chaos: momentum correlations are lost between collisions.

\item separation of time-scales: the evolution of the system is slow compared with the microscopic QFT processes.

\end{enumerate}
The former of these two assumptions enables us to express the correlation functions (within the collision integrals) in terms of (products of) single-particle distribution functions, viz.~the number density and pair correlations. In order to make use of the latter assumption, and closely following Ref.~\cite{Dev:2014laa}, we consider the integral
\begin{equation}\label{I WW 1}
 \mathcal{I}\ =\ \int_{t_{0}}^{t}\sd{d}t'\;\text{Tr}\,\big\{\big[\hat{F}(t),\hat{H}_{\text{int}}(t')\big]\hat{\rho}(t')\big\}\;,\qquad \hat{F}(t)\ \coloneqq\ \big[\hat{N}_{\mathbf{k}}(t),\hat{H}_{\text{int}}(t)\big]\;.
\end{equation}
Inserting the Fourier transforms of both $\hat{F}(t)$ and $\hat{H}_{\text{int}}(t')$, we have
\begin{equation}
 \mathcal{I}\ =\ \int_{t_{0}}^{t}\,\sd{d}t'\int_{-\infty}^{+\infty}\frac{\sd{d}\omega}{2\pi}\int_{-\infty}^{+\infty}\frac{\sd{d}\omega'}{2\pi}\,e^{i\omega t}e^{i\omega' t'}\,\text{Tr}\,\big\{\big[\hat{\tilde{F}}(\omega),\hat{\tilde{H}}_{\text{int}}(\omega')\big]\hat{\rho}(t')\big\}\;,
\end{equation}
and after making a change of variables $\omega\to \omega-\omega'$, we can re-express this as 
\begin{equation}\label{I WW 2}
 \mathcal{I}\ =\ \int_{t_{0}}^{t}\sd{d}t'\int_{-\infty}^{+\infty}\frac{\sd{d}\omega}{2\pi}\int_{-\infty}^{+\infty}\frac{\sd{d}\omega'}{2\pi}\;e^{i\omega t}\,e^{i\omega'(t'-t)}\,\text{Tr}\,\big\{\big[\hat{\tilde{F}}(\omega-\omega'),\hat{\tilde{H}}_{\text{int}}(\omega')\big]\hat{\rho}(t')\big\}\;.
\end{equation}
The exponential \smash{$e^{i\omega'\left(t'-t\right)}$} oscillates rapidly when $t\nsim t'$, and, as long as \smash{$\hat{\tilde{F}}(\omega-\omega')$} is not strongly peaked, the integral over $\omega'$ is dominated by $t\sim t'$. We can therefore replace $\hat{\rho}(t')$ by $\hat{\rho}(t)$ and extend the lower limit of the $t'$ integral to $-\infty$. Subsequently making the judicious change of variables $t' \to t'-t$, we  obtain
\begin{equation}\label{I WW 4}
 \mathcal{I}\ \simeq\ \int_{-\infty}^{0}\sd{d}t'\int_{-\infty}^{+\infty}\frac{\sd{d}\omega}{2\pi}\int_{-\infty}^{+\infty}\frac{\sd{d}\omega'}{2\pi}\;e^{i\omega t}\,e^{i\omega't'}\,\text{Tr}\,\big\{\big[\hat{\tilde{F}}(\omega-\omega'),\hat{\tilde{H}}_{\text{int}}(\omega')\big]\hat{\rho}(t)\big\}\;.
\end{equation}
We can now rewrite the integral over $t'$ by means of the identity
\begin{equation}
 \int_{-\infty}^{0}\sd{d}t'\;e^{i\omega't'}\ =\ \frac{1}{2}\int_{-\infty}^{+\infty}\sd{d}t'\;e^{i\omega't'}\:-\:i\mathcal{P}\,\frac{1}{\omega'}\;,
\end{equation}
where $\mathcal{P}$ denotes the Cauchy principal value. Doing so, we find
\begin{flalign}\label{I WW 5}
 \mathcal{I}\ & \simeq\ \frac{1}{2}\int_{-\infty}^{+\infty}\sd{d}t'\;\text{Tr}\,\big\{\big[\big[\hat{N}_{\mathbf{k}}(t),\hat{H}_{\text{int}}(t)\big],\hat{H}_{\text{int}}(t')\big]\hat{\rho}(t)\big\}\nonumber\\ &\qquad-\:i\mathcal{P}\int_{-\infty}^{+\infty}\frac{\sd{d}\omega'}{2\pi}\;\frac{e^{i\omega't'}}{\omega'}\,\text{Tr}\,\big\{\big[\big[\hat{N}_{\mathbf{k}}(t),\hat{H}_{\text{int}}(t)\big],\hat{\tilde{H}}_{\text{int}}(\omega')\big]\hat{\rho}(t)\big\}\;.
\end{flalign}
The collision terms arise from the first term in Eq.~\eqref{I WW 5}; the second term gives rise to dispersive corrections, which we hereafter neglect.

Implementing the above approximations, we arrive at the following coupled set of Markovian master equations for the number density $N_{\mathbf{k}}(t)$, and the pair correlations $M_{\mathbf{k}}(t)$ and $M^{\ast}_{\mathbf{k}}(t)$, valid at order $\lambda^2$:
\begin{subequations}
\label{eq:fullMarkmaster}
\begin{flalign}
 \dot{N}_{\mathbf{k}}(t)\ &\simeq\ \frac{\dot{\omega}_{\mathbf{k}}(t)}{\omega_{\mathbf{k}}(t)}\,\text{Re}\, M_{\mathbf{k}}(t)\:-\:i\big\langle\big[\hat{N}_{\mathbf{k}}(t),\hat{H}_{\text{int}}(t)\big]\big\rangle_{t_{0}}\nonumber\\&\qquad-\:\frac{1}{2}\int_{-\infty}^{\infty}\mathrm{d}t'\;\big\langle\big[\big[\hat{N}_{\mathbf{k}}(t),\hat{H}_{\text{int}}(t)\big],\hat{H}_{\text{int}}(t')\big]\big\rangle_{t}\;,\label{final ww num density evo}\displaybreak\\[0.5em] \dot{M}_{\mathbf{k}}(t)\ &\simeq\ -\:2i\omega_{\mathbf{k}}(t)M_{\mathbf{k}}(t)\:+\:\frac{1}{2}\,\frac{\dot{\omega}_{\mathbf{k}}(t)}{\omega_{\mathbf{k}}(t)}\,\big(2N_{\mathbf{k}}(t)\:+\:1\big)\:\nonumber\\ &\qquad -\:i\big\langle\big[\hat{M}_{\mathbf{k}}(t),\hat{H}_{\text{int}}(t)\big]\big\rangle_{t_{0}}\:-\:\frac{1}{2}\int_{-\infty}^{\infty}\mathrm{d}t'\;\big\langle\big[\big[\hat{M}_{\mathbf{k}}(t),\hat{H}_{\text{int}}(t)\big],\hat{H}_{\text{int}}(t')\big]\big\rangle_{t}\;,\label{final ww M evo}\\[0.5em] \dot{M}^{\ast}_{\mathbf{k}}(t)\ &\simeq\ +\:2i\omega_{\mathbf{k}}(t)M^{\ast}_{\mathbf{k}}(t)\:+\:\frac{1}{2}\,\frac{\dot{\omega}_{\mathbf{k}}(t)}{\omega_{\mathbf{k}}(t)}\,\big(2N_{\mathbf{k}}(t)\:+\:1\big)\nonumber\\&\qquad -\:i\big\langle\big[\hat{M}^{\dagger}_{\mathbf{k}}(t),\hat{H}_{\text{int}}(t)\big]\big\rangle_{t_{0}}\:-\:\frac{1}{2}\int_{-\infty}^{\infty}\mathrm{d}t'\;\big\langle\big[\big[\hat{M}^{\dagger}_{\mathbf{k}}(t),\hat{H}_{\text{int}}(t)\big],\hat{H}_{\text{int}}(t')\big]\big\rangle_{t}\;.\label{final ww M* evo}
\end{flalign} 
\end{subequations}
While we focus here on the context of preheating, these Markovian master equations can be applied more generally to describe the evolution of any interacting scalar field with a time-dependent mass term. We emphasize again that the particle production terms in the master equation for the number density, Eq.~\eqref{final ww num density evo}, are proportional to \smash{$M_{\mathbf{k}}(t)$} and \smash{$M^{\ast}_{\mathbf{k}}(t)$}. We see, therefore, that non-adiabatic particle production depends crucially on the existence and non-vanishing values of pair correlations, such that any processes that destroy these coherences will suppress (and eventually shut off) the resonant production (cf.~Ref.~\cite{Morikawa:1984dz}).\footnote{Note that, in a realistic scenario, the inflaton condensate will decay. Once it has completely diminished, the production terms in the master equations will vanish and the system will continue to thermalize. At this point, the evolution of the system will become adiabatic, and the pair correlations \smash{$M_{\mathbf{k}}(t)$} and \smash{$M^{\ast}_{\mathbf{k}}(t)$} will decohere.} We further note from Eq.~\eqref{eq:fullMarkmaster} that, since $M_{\mathbf{k}}(t)$ and $M^{\ast}_{\mathbf{k}}(t)$ source $N_{\mathbf{k}}(t)$, one should expect their values to be of the same order of magnitude during the preheating phase, so long as \smash{$\lvert\dot{\omega}_{\mathbf{k}}/\omega^{2}_{\mathbf{k}}\rvert\,\sim\,\mathcal{O}(1)$}, which is true during intervals of non-adiabaticity.


\subsection{Collision terms}\label{ssec:collisions}

We now wish to express the master equations entirely in terms of $N_{\mathbf{k}}(t)$, $M_{\mathbf{k}}(t)$ and $M^{\ast}_{\mathbf{k}}(t)$, so as to obtain a self-consistent set of evolution equations. In order to do so, we first need to solve the Heisenberg equations for $\hat{a}_{\mathbf{k}}(t)$ and $\hat{a}^{\dagger}_{\mathbf{k}}(t)$ (see Eq.~\eqref{time evo aa*}), so that we can evolve all of the operators appearing in the collision terms to equal times, namely the time $t$ of the density operator.

To this end, we assume a Bogoliubov ansatz of the form given by Eq.~\eqref{aa* ansatz}, writing the field at time $t'\leq t$ as
\begin{equation}
 \hat{\chi}(t',\mathbf{x})\ =\ \int\!\frac{\sd{d}^3\mathbf{k}}{(2\pi)^{3}}\,\Big[\tilde{\chi}_{\mathbf{k}}(t',t)\hat{a}_{\mathbf{k}}(t)e^{+i\mathbf{k}\cdot\mathbf{x}}\:+\:\tilde{\chi}^{\ast}_{\mathbf{k}}(t',t)\hat{a}^{\dagger}_{\mathbf{k}}(t)e^{-i\mathbf{k}\cdot\mathbf{x}}\Big]\;,
\end{equation}
where the mode function $\tilde{\chi}_{\mathbf{k}}(t',t)$ is given by
\begin{equation}\label{mode functions}
 \tilde{\chi}_{\mathbf{k}}(t',t)\ \coloneqq\ \frac{1}{\sqrt{2\omega_{\mathbf{k}}(t')}}\,\big[\alpha_{\mathbf{k}}(t',t)\:+\:\beta_{\mathbf{k}}(t',t)\big]\;.
\end{equation}
Its evolution can be cast in the form a Mathieu equation, as in Eq.~\eqref{mathieu}. Since Bogoliubov transformations preserve the canonical algebra, the functions $\alpha_{\mathbf{k}}(t',t)$ and $\beta_{\mathbf{k}}(t',t)$ satisfy the constraint
\begin{equation}
 \big\lvert\alpha_{\mathbf{k}}(t',t)\big\rvert^{2}\:-\:\big\lvert\beta_{\mathbf{k}}(t',t)\big\rvert^{2}\ =\ 1\;,
\end{equation}
and we can invert Eq.~\eqref{aa* ansatz} in order to relate the creation and annihilations operators at times $t'\leq t$ to those at time $t$:
\begin{subequations}\label{ab inverse}
\begin{flalign}
 \hat{a}_{\mathbf{k}}(t')\ &=\ \alpha_{\mathbf{k}}(t',t)\hat{a}_{\mathbf{k}}(t)\:+\:\beta^{\ast}_{\mathbf{k}}(t',t)\hat{a}^{\dagger}_{-\mathbf{k}}(t)\;,\\ \hat{a}^{\dagger}_{\mathbf{k}}(t')\ &=\ \alpha_{\mathbf{k}}^{\ast}(t',t)\hat{a}^{\dagger}_{\mathbf{k}}(t)\:+\:\beta_{\mathbf{k}}(t',t)\hat{a}_{-\mathbf{k}}(t)\;.
\end{flalign}
\end{subequations}
Moreover, we can show that
\begin{subequations}
\begin{flalign}
 \alpha_{\mathbf{k}}(t,t'')\ &=\ \alpha_{\mathbf{k}}(t,t')\alpha_{\mathbf{k}}(t',t'')\:+\:\beta_{\mathbf{k}}^{\ast}(t,t')\beta_{\mathbf{k}}(t',t'')\;,\\  \beta_{\mathbf{k}}(t,t'')\ &=\ \beta_{\mathbf{k}}(t,t')\alpha_{\mathbf{k}}(t',t'')\:+\:\alpha^{\ast}_{\mathbf{k}}(t,t')\beta_{\mathbf{k}}(t',t'')\;.
\end{flalign}
\end{subequations}

We now recall the Wigner-Weisskopf approximation of the collision terms, described earlier in subsection~\ref{ssec:master eqs}, wherein we argued that the collision integral is dominated by times $t'\sim t$. This allowed us to replace $\hat{\rho}(t')$ by $\hat{\rho}(t)$ and extend the upper and lower limits of the integration over $t'$ to positive and negative infinity, respectively. Once all of the creation and annihilation operators have been evolved to the time $t$, we are therefore interested in the Bogoliubov coefficients only for times $t'$ near $t$, while treating the integration over this neighbourhood in $t'$ as effectively infinite as far as the dynamics of the fast modes is concerned. It follows from Eq.~\eqref{ab inverse} that $\beta_{\mathbf{k}}$ vanishes for $t'=t$, and we therefore make use of the following approximate solutions in the collision integral (see Appendix~\ref{sec:appendix b}):
\begin{subequations}
\label{ab sols}
\begin{align}
\label{eq:asol}
 \alpha_{\mathbf{k}}(t',t)\ &\simeq\ e^{+i\bar{\omega}_{\mathbf{k}}(t-t')}\;,\\
  \beta_{\mathbf{k}}(t',t)\ &\simeq\ 0\;,
\end{align}
\end{subequations}
where
\begin{equation}\label{time-average}
  \bar{\omega}_{\mathbf{k}} \ =\ \left(1-\frac{g\varphi_{0}^{2}}{8\omega^{2}_{\mathbf{k}}\vert_{\text{max}}}\right)\omega_{\mathbf{k}}\vert_{\text{max}}\;,\qquad \omega_{\mathbf{k}}\vert_{\text{max}}\ \coloneqq\ \sqrt{\mathbf{k}^{2}+m_{\chi}^{2}+g\varphi_{0}^{2}/2}\;,
\end{equation}
is (approximately) the time-averaged energy.

The above approximation imposes adiabatic evolution in the collision integral between the times $t$ and $t'$. Since the periods of non-adiabatic particle production are much shorter than the intervening periods of adiabatic evolution, this is expected to be a valid procedure, so long as the time-scale for the collisions is much larger than that of each burst of particle production. For broad resonance, the production rate is \smash{$\Gamma_*\coloneqq 1/\Delta t_{\ast}\sim(\sqrt{g/2}m_{\phi}\varphi_{0})^{1/2}$}, and the produced particles typically have momenta lying in the interval \smash{$0\leq\lvert \mathbf{k}\rvert\lesssim m_{\phi}(q/4)^{1/4}$} \smash{$\approx 4m_{\phi}< \sqrt{g/2}\varphi_0$} for  $q= \mathcal{O}
(10^3)$~\cite{Kofman:1997yn}. During the very early stages of preheating, we can therefore estimate the total number density to be \smash{$N\propto m^{3}_{\phi}$}. The collision rate is given by $\Gamma(\chi\chi\rightarrow\chi\chi) \sim \lvert\mathbf{v}_{\text{rel}}\rvert N\sigma(\chi\chi\to\chi\chi)$. The relative velocity scales as $\lvert\mathbf{v}_{\text{rel}}\rvert\,\sim\,2\lvert\mathbf{k}\rvert/m_{\text{eff}}(t)$, and herein we assume that this is of order one to obtain a conservative estimate for the collision rate.~When $m_{\rm eff}^2\simeq g\varphi_0^2/2$ is maximal, $\lvert\mathbf{k}\rvert\lesssim m_\text{eff}$, and the two-to-two scattering cross-section scales as \smash{$\sigma(\chi\chi\rightarrow\chi\chi)\sim\lambda^{2}/(64\pi\,g\varphi^{2}_{0})$}. On the other hand, when $\varphi$ nears the turning point of an oscillation and for modes with $|\mathbf{k}|\lesssim m_{\chi}$, the cross-section scales as \smash{$\sigma(\chi\chi\rightarrow\chi\chi)\sim\lambda^{2}/(128\pi m_{\chi}^2)$}. For modes with $\lvert\mathbf{k}\rvert\lesssim m_{\phi}$, the cross-section scales as \smash{$\sigma(\chi\chi\rightarrow\chi\chi)\sim\lambda^{2}/(128\pi m_{\phi}^2)$}. Hence, in order to achieve the separation of scales required above,~i.e.~$\Gamma(\chi\chi\to\chi\chi)\ll\Gamma_*$, we need \mbox{\smash{$\lambda^{2}m_{\phi}^{5/2}(\sqrt{g/2}\varphi_0)^{-1/2}\{1/(g\varphi_0^2/2),1/m_{\chi}^2,1/m_{\phi}^2\}/(128\pi)$}} $\ll 1$, and all three cases can be satisfied for $\lambda = \mathcal{O}(0.1)$. 

At this point, we note that the particle number can become extremely large during preheating, in which case the system can become effectively strongly coupled~\cite{Son:1996uv}, such that many-to-many processes dominate over the two-to-two scatterings that we consider here. However, we are interested only in the first few oscillations of the inflaton, wherein the particle number remains relatively small and the collision integral amounts to perturbatively small corrections to the Boltzmann equations. Nevertheless, one would anticipate that such strong coupling would only increase the impact of the collisions on the dynamics of the resonance.

We can now proceed to evaluate the remaining expectation values in the Markovian master equations [Eq.~\eqref{eq:fullMarkmaster}]. (We refer the reader to Appendix \ref{sec:appendix c} for all the gory details.) In order to do so, we make two additional assumptions about the system: (i) that the state is (approximately) Gaussian so that all higher-order correlations can be expressed in terms of one- and two-point functions by Wick's theorem and (ii) that the system is spatially homogeneous, such that momentum-space two-point correlation functions can be written in the form
\begin{subequations}
\begin{flalign}
 \big\langle\hat{\mathcal{O}}_{\mathbf{p}}(t)\hat{\mathcal{O}}_{\mathbf{k}}(t)\big\rangle\ &=\ \frac{(2\pi)^{3}}{\text{Vol}}\,\delta^{(3)}(\mathbf{p}+\mathbf{k})\big\langle\hat{\mathcal{O}}_{-\mathbf{k}}(t)\hat{\mathcal{O}}_{\mathbf{k}}(t)\big\rangle\;,\\ \big\langle\hat{\mathcal{O}}^{\dagger}_{\mathbf{p}}(t)\hat{\mathcal{O}}_{\mathbf{k}}(t)\big\rangle \ &=\ \frac{(2\pi)^{3}}{\text{Vol}}\,\delta^{(3)}(\mathbf{p}-\mathbf{k})\big\langle\hat{\mathcal{O}}_{\mathbf{k}}(t)\hat{\mathcal{O}}_{\mathbf{k}}(t)\big\rangle\;.
\end{flalign}
\end{subequations}
Moreover, we are interested only in the collision terms, arising at $\mathcal{O}(\lambda^2)$, that drive the thermalization. We neglect the radiative corrections to the $\chi$ mass, which arise from both the $\mathcal{O}(\lambda)$ and $\mathcal{O}(\lambda^2)$ terms in Eq.~\eqref{eq:fullMarkmaster}.

The final Boltzmann equations are as follows:
\begin{subequations}
\label{final eoms 1}
\begin{align}
 \dot{N}_{\mathbf{k}}(t)\ &\simeq\ \frac{\dot{\omega}_{\mathbf{k}}(t)}{\omega_{\mathbf{k}}(t)}\,{\rm Re}\, M_{\mathbf{k}}(t)\:+\:{\rm Re}\,\mathcal{C}^{(N)}_{\mathbf{k}}[N,M;t]\;,\label{n eom}\\[0.5em] {\rm Re}\,\dot{M}_{\mathbf{k}}(t)\ &\simeq\ +\:2\omega_{\mathbf{k}}(t)\,{\rm Im}\,M_{\mathbf{k}}(t)\:+\:\frac{1}{2}\,\frac{\dot{\omega}_{\mathbf{k}}(t)}{\omega_{\mathbf{k}}(t)}\,\Big(2N_{\mathbf{k}}(t)\:+\:1\Big)\:+\:{\rm Re}\,\mathcal{C}^{(M)}_{\mathbf{k}}[N,M;t]\;,\label{M eom}\\[0.75em]
{\rm Im}\,\dot{M}_{\mathbf{k}}(t)\ &\simeq\ -\:2\omega_{\mathbf{k}}(t)\,{\rm Re}\,M_{\mathbf{k}}(t)\:+\:{\rm Im}\,\mathcal{C}^{(M)}_{\mathbf{k}}[N,M;t]\;,\label{Mstar eom}
\end{align}
\end{subequations}
where the $\mathcal{C}$'s are the collision terms. The latter can be written in the form 
\begin{equation}
\mathcal{C}_{\mathbf{k}}^{(\mathcal{O})}[N,M,t]\ =\ \frac{1}{2}\int_{-\infty}^{\infty}\mathrm{d}t'\int_{\mathbf{x},\mathbf{y}}\Big[\Pi^{<}(x,y)G_{\mathcal{O}_{\mathbf{k}}}^{>}(y,x)\:-\:\Pi^{>}(x,y)G_{\mathcal{O}_{\mathbf{k}}}^{<}(y,x)\Big]\label{n collisions}\;,
\end{equation}
where $\mathcal{O}\in\lbrace N,\,M\rbrace$, $x^{\mu}=(t,\mathbf{x})$ and $y^{\mu}=(t',\mathbf{y})$, with $t'<t$. In addition, we have defined 
\begin{subequations}\label{wightman}
\begin{align}
  G_{\mathcal{O}_{\mathbf{k}}}^{>}(x,y)\ &\coloneqq\ \big\langle\hat{\chi}(t,\mathbf{x})\big[\hat{\mathcal{O}}_{\mathbf{k}}(t),\hat{\chi}(t',\mathbf{y})\big]\big\rangle_{t}\;,\\[0.8em]
G_{\mathcal{O}_{\mathbf{k}}}^{<}(x,y)\ &\coloneqq\ \big\langle\big[\hat{\mathcal{O}}_{\mathbf{k}}(t),\hat{\chi}(t',\mathbf{y})\big]\hat{\chi}(t,\mathbf{x})\big\rangle_{t}\;,\\[0.2em] 
 \Pi^{>}(x,y)\ &\coloneqq\ \frac{\lambda^{2}}{3!}\,\big\langle\hat{\chi}(t,\mathbf{x})\hat{\chi}(t',\mathbf{y})\big\rangle_{t}^{3}\;,\\
\Pi^{<}(x,y)\ &\coloneqq\ \frac{\lambda^{2}}{3!}\,\big\langle\hat{\chi}(t',\mathbf{y})\hat{\chi}(t,\mathbf{x})\big\rangle_{t}^{3}\;.
\end{align}
\end{subequations}
The \smash{$G_{\mathcal{O}_{\mathbf{k}}}^{\lessgtr}(x,y)$} are related to the positive- ($>$) and negative-frequency ($<$) Wightman propagators and the $\Pi^{\lessgtr}(x,y)$ correspond to the two cut self-energies.

\begin{figure}[t!]

\begin{center}
 
 \subfloat[\label{fig:sunseta}]{\includegraphics[scale=0.6]{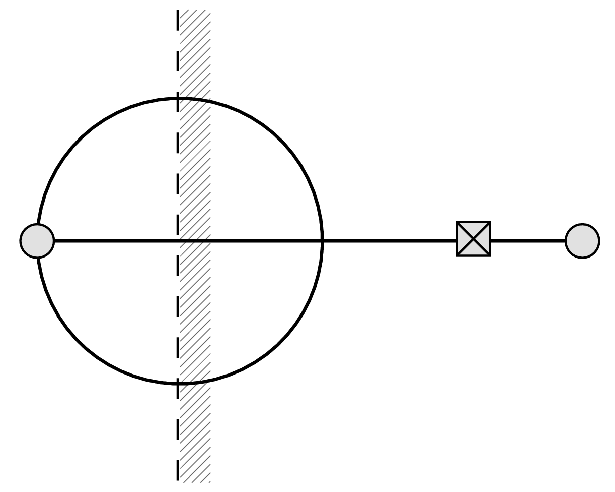}} \qquad
 \subfloat[\label{fig:sunsetb}]{\raisebox{1.3mm}{{\includegraphics[scale=0.6, trim={5cm 10cm 5cm 9cm}, clip]{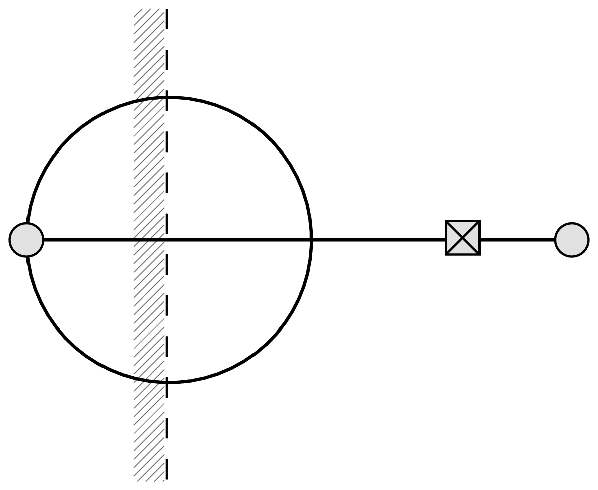}}}}

\end{center}

\caption{The forward (a) and backward (b) cuts of the scalar two-loop ``sunset'' diagram. The small shaded circles indicate coincident points, and the crossed boxes indicate insertions of the operator $\hat{\mathcal{O}}_{\mathbf{k}}$. The net energy flow is from the unshaded to shaded regions.}\label{fig:sunset}

\end{figure}

While we calculate the collision terms directly from the operator algebra, we have written them in terms of the self-energies and Green's functions so as to make contact with approaches based on non-equilibrium quantum field theory (see, e.g., Refs.~\cite{Blaizot:2001nr,Berges:2004yj}). In this case, the collision terms can be associated with the absorptive cuts of the non-equilibrium self-energies, which can be calculated by means of the Kobes-Semenoff~\cite{Kobes:1985kc,Kobes:1986za} cutting rules that generalize those of Cutkosky~\cite{Cutkosky:1960sp}, and 't Hooft and Veltman~\cite{tHooft:1973wag}. The gain and loss terms are then associated with the forward and backward cuts of the two-loop sunset diagram, as depicted in Fig.~\ref{fig:sunset}, wherein all cut lines are placed on-shell with the net energy flow proceeding from the unshaded to the shaded regions.

\begin{figure}[t!]

\begin{center}
 \subfloat[]{\raisebox{-22mm}{\includegraphics[scale=0.8, trim={3cm 21cm 10cm 2cm}, clip]{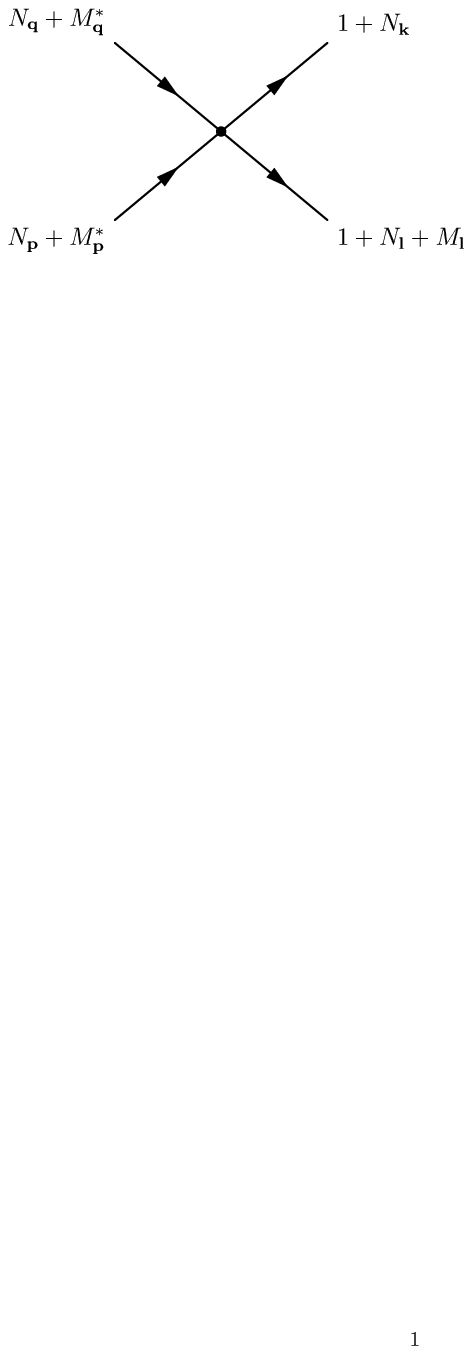}}$\quad-\quad$\raisebox{-22mm}{\includegraphics[scale=0.8, trim={3cm 21cm 9cm 2cm}, clip]{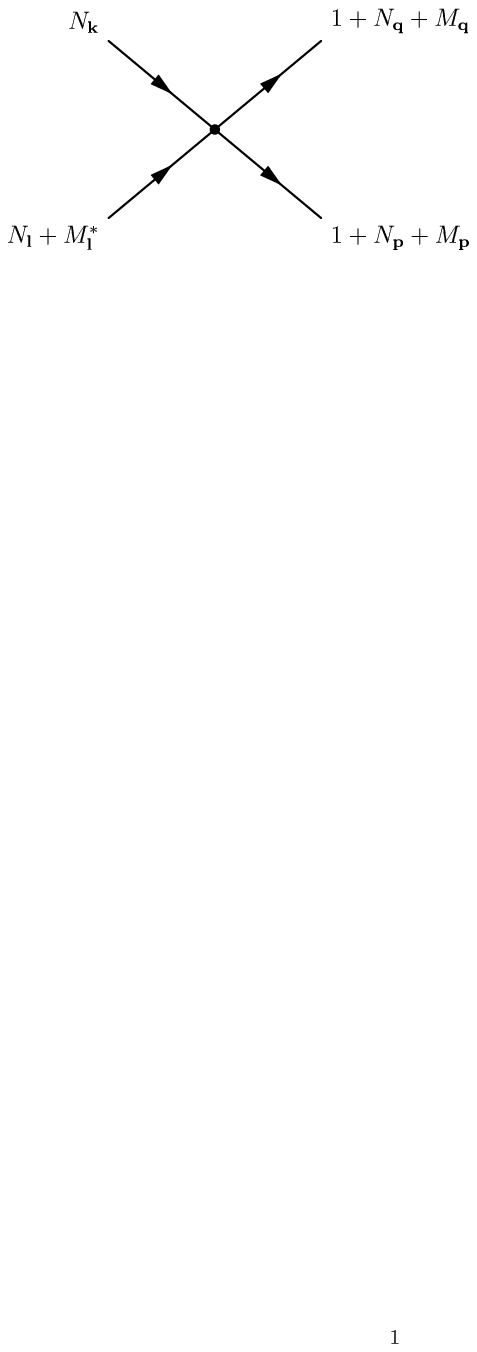}}} \vfill\vfill\subfloat[]{\raisebox{-22mm}{\includegraphics[scale=0.8, trim={3cm 21cm 10cm 2cm}, clip]{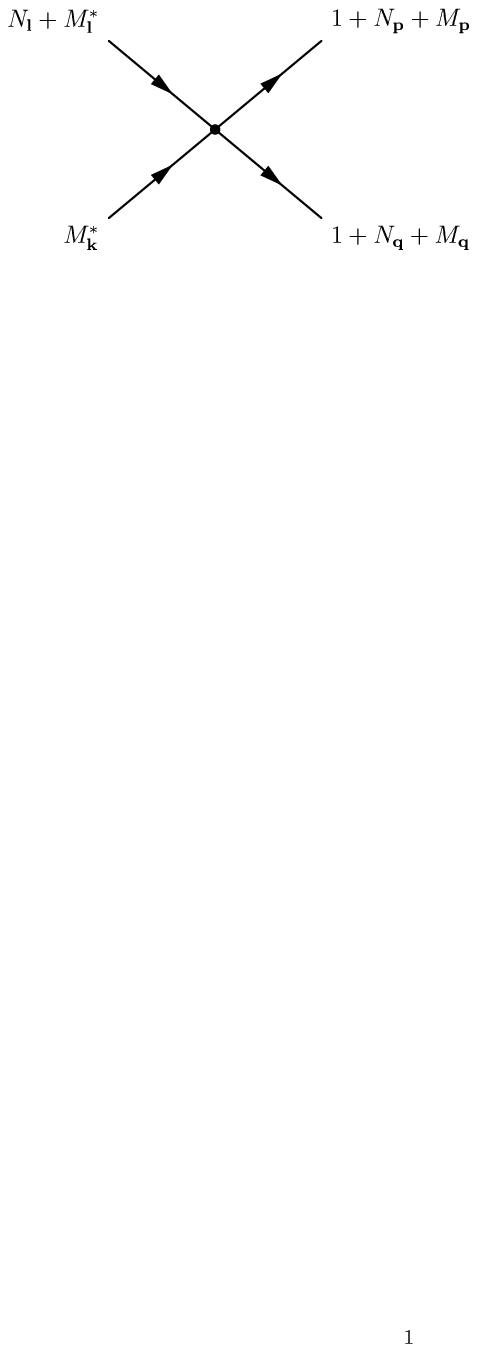}}$\quad-\quad$\raisebox{-22mm}{\includegraphics[scale=0.8, trim={3cm 21cm 10cm 2cm}, clip]{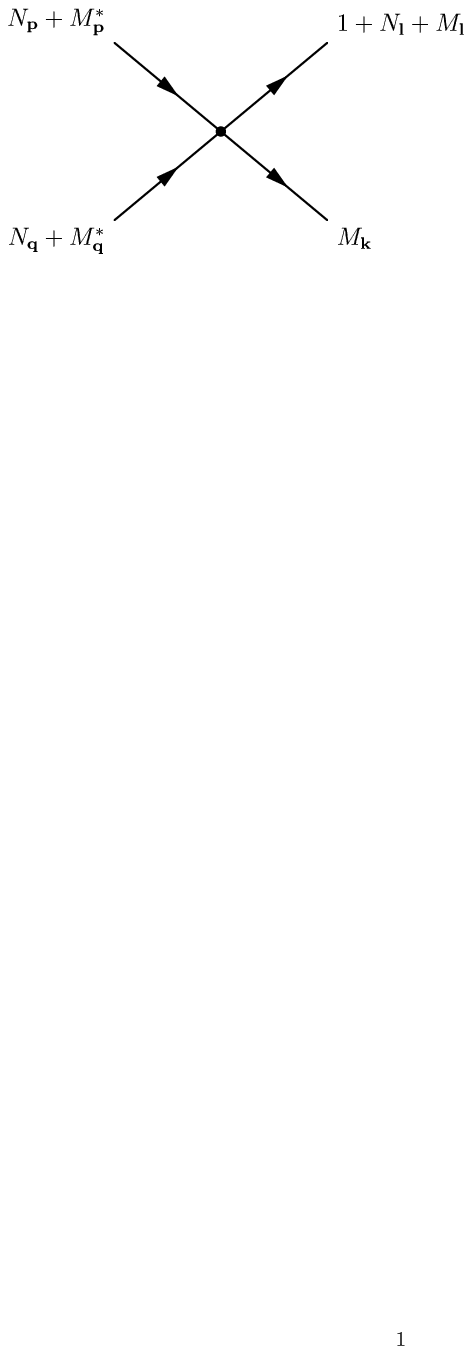}}}

\end{center}

\caption{Feynman diagrams of the two-to-two scattering processes in the number density collision integral \smash{$\mathcal{C}_{\mathbf{k}}^{(N)}[N,M,t]$}: (a) gain and loss terms, where the ``external'' momentum $\mathbf{k}$ is associated with the number density; (b) gain and loss terms, where the ``external'' momentum $\mathbf{k}$ is associated with the (complex conjugate of the) pair correlation. Note that the momentum $\mathbf{l}=\mathbf{p}+\mathbf{q}-\mathbf{k}$ is determined by three-momentum conservation at each vertex.}\label{fig:scattering}

\end{figure}

The Wigner-Weisskopf approximation enforces quasi-energy-momentum conservation\footnote{The ``quasi-energy'' conservation refers to the fact that it is not the instantaneous time-dependent energies that are conserved but rather their approximate time averages.} at each interaction vertex. As a result, the only kinematically allowed collisional processes are two-to-two scatterings. The collision terms then reduce to
\begin{equation}
  \mathcal{C}_{\mathbf{k}}^{(\mathcal{O})}[N,M,t]\ \simeq\ \frac{\lambda^{2}}{2}\,\int {\rm d}\Pi_{\mathbf{p},\mathbf{q},\mathbf{k}}\;f^{(\mathcal{O})}_{\mathbf{p},\mathbf{q},\mathbf{k}}[N,M;t]\label{final n C}\;,
\end{equation}
where
\begin{equation}
 {\rm d}\Pi_{\mathbf{p},\mathbf{q},\mathbf{k}}\ =\ \frac{\sd{d}^{3}\mathbf{p}}{(2\pi)^{3}}\,\frac{\sd{d}^{3}\mathbf{q}}{(2\pi)^{3}}\,2\pi\delta(\bar{\omega}_{\mathbf{k}}+\bar{\omega}_{\mathbf{p}+\mathbf{q}-\mathbf{k}}-\bar{\omega}_{\mathbf{p}}-\bar{\omega}_{\mathbf{q}})\,\prod_{\boldsymbol\kappa}\frac{1}{2\bar{\omega}_{\boldsymbol\kappa}}\;,\quad \boldsymbol\kappa\in\lbrace\mathbf{k},\,\mathbf{p},\,\mathbf{q},\,(\mathbf{p}+\mathbf{q}-\mathbf{k})\rbrace\;.\label{phase space}
\end{equation}
Notice that the time-dependent energies in the phase-space measures (cf.~Eq.~\eqref{eq:measure}) have been replaced by $\bar{\omega}_{\boldsymbol\kappa}$ so as to be consistent with the approximate solution for $\alpha_{\mathbf{k}}(t',t)$ in Eq.~\eqref{eq:asol}. The functions \smash{$f^{(\mathcal{O})}_{\mathbf{p},\mathbf{q},\mathbf{k}}[N,M;t]$} contain the statistical weights and are given in full in Appendix~\ref{sec:appendix c}. Their real and imaginary parts are given by
\begin{subequations}\label{stat weight}
\begin{flalign}
 \mathrm{Re}f^{(N)}_{\mathbf{p},\mathbf{q},\mathbf{k}}[N,M;t]\ &=\ -\:\mathrm{Re}f^{(M)}_{\mathbf{p},\mathbf{q},\mathbf{k}}[N,M;t]\nonumber\\ &=\ \mathrm{Re}\Big[\!\left(1+N_{\mathbf{k}}-M_{\mathbf{k}}\right)\left(N_{\mathbf{p}}+M^{\ast}_{\mathbf{p}}\right)\left(N_{\mathbf{q}}+M^{\ast}_{\mathbf{q}}\right)\left(1+N_{\mathbf{p}+\mathbf{q}-\mathbf{k}}+M_{\mathbf{p}+\mathbf{q}-\mathbf{k}}\right) \nonumber\\
&\quad -\:\left(N_{\mathbf{k}}-M^{\ast}_{\mathbf{k}}\right)\left(1+N_{\mathbf{p}}+M_{\mathbf{p}}\right)\left(1+N_{\mathbf{q}}+M_{\mathbf{q}}\right)\left(N_{\mathbf{p}+\mathbf{q}-\mathbf{k}}+M^{\ast}_{\mathbf{p}+\mathbf{q}-\mathbf{k}}\right)\!\Big]\;,\\ 
\mathrm{Im}f^{(M)}_{\mathbf{p},\mathbf{q},\mathbf{k}}[N,M;t]\ &=\ \mathrm{Im}\Big[\!\left(N_{\mathbf{k}}+M^{\ast}_{\mathbf{k}}\right)\left(1+N_{\mathbf{p}}+M_{\mathbf{p}}\right)\left(1+N_{\mathbf{q}}+M_{\mathbf{q}}\right)\left(N_{\mathbf{p}+\mathbf{q}-\mathbf{k}}+M^{\ast}_{\mathbf{p}+\mathbf{q}-\mathbf{k}}\right) \nonumber\\
&\quad -\:\left(1+N_{\mathbf{k}}+M_{\mathbf{k}}\right)\left(N_{\mathbf{p}}+M^{\ast}_{\mathbf{p}}\right)\left(N_{\mathbf{q}}+M^{\ast}_{\mathbf{q}}\right)\left(1+N_{\mathbf{p}+\mathbf{q}-\mathbf{k}}+M_{\mathbf{p}+\mathbf{q}-\mathbf{k}}\right)\!\Big] \;,
\end{flalign}
\end{subequations}
comprising the gain and loss terms that arise from two-to-two scattering processes. These are shown diagrammatically in Fig.~\ref{fig:scattering}, where we have treated the contributions from \smash{$N_{\mathbf{k}}$} and \smash{$M_{\mathbf{k}}^{(*)}$} separately. We draw attention to the relative signs between the $N$'s and $M$'s in the factors corresponding to the external momentum $\mathbf{k}$ in Eq.~\eqref{stat weight}, which do not occur for the internal statistical factors. These have arisen because the gain and loss terms are interchanged in the contributions from the external pair correlation $M_{\mathbf{k}}$ relative to those arising from the number density $N_{\mathbf{k}}$.

\section{Numerical example}
\label{sec:numerics}

In this section, we present numerical solutions to the Boltzmann equations in Eq.~\eqref{final eoms 1} for the very early stages of preheating. Throughout this analysis, we have focused on the case of broad resonance (i.e.~that which occurs over a broad range of momenta and corresponds to the condition $q=(g\varphi_{0}^{2})/(8m_{\phi}^{2})\gg 1$), since reheating becomes extremely efficient, thereby enabling a relatively large occupancy for each momentum mode to build up in just a few oscillations of the inflaton field. In this case, one expects the effect of the collision terms to be more pronounced at these early stages compared to the regime of narrow resonance. With this in mind, we choose the model parameters as follows: $\varphi_{0}=10^{5}\,m_{\phi}$, $m_{\chi}=m_{\phi}/10$ and $g=5\times 10^{-7}$, such that $q\sim\mathcal{O}(10^{3})$. 

The results that we present were obtained by means of a fourth-order Runge-Kutta differential solver implemented in Mathematica and involving five non-trivial phase-space integrals over the magnitudes of the momenta $\mathbf{p}$ and $\mathbf{q}$, the relative angle between them and the relative angles of one of these momenta to the external momentum $\mathbf{k}$. We remark that the approximations made in order to reduce the solutions for $\alpha_{\mathbf{k}}(t,t')$ and $\beta_{\mathbf{k}}(t,t')$ (cf.~Eq.~\eqref{ab sols}) to a form yielding (quasi-)energy-conserving Dirac delta functions introduce an error of at most $\sim 15\%$ to the collision integrals (see Appendix~\ref{sec:appendix b}). This is, however, anticipated to be a global error, rather than a relative error between the contributions to each collision integral, and therefore is expected to have little impact on the inferences that follow.

\begin{figure}[t!]

\vspace{3em}

\begin{center}

\subfloat[\label{fig:free evo}]{\includegraphics[width=0.76\textwidth]{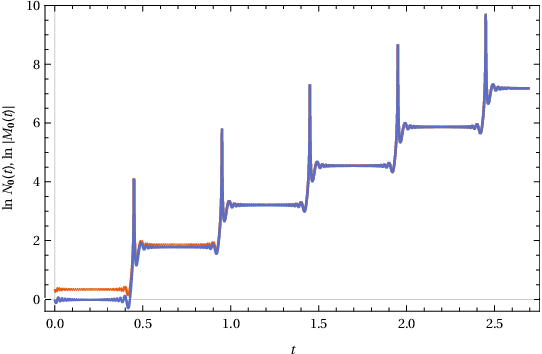}}\vfill\vfill\hspace{0.25mm}
\subfloat[\label{fig:free n M evo}]{\includegraphics[width=0.75\textwidth]{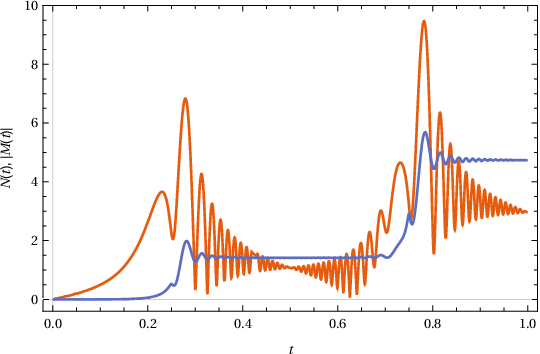}}

\end{center}

\caption{(a) Time evolution (measured in units of $m_{\phi}/2\pi$) of the $\chi$ particle number density $N_{\mathbf{k}}$ (blue) and pair correlation $|M_{\mathbf{k}}|$ (orange) for the mode $\mathbf{k}=\mathbf{0}$ in the regime of broad resonance ($q\sim 10^{3}$) for the collisionless case $\lambda =0$. (b) Evolution of the integrated number density (blue) and pair correlation (orange), i.e.~$N(t)=\int\!\frac{{\rm d}^3\mathbf{k}}{(2\pi)^3}\,N_{\mathbf{k}}$ and $|M(t)|=\int\!\frac{{\rm d}^3\mathbf{k}}{(2\pi)^3}\,|M_{\mathbf{k}}|$ up to $t\ =\ 2\pi/m_{\phi}$.}\label{fig:lambda zero}

\end{figure}

In Fig.~\ref{fig:free evo}, we show the evolution of the (natural) logarithm of the number density and the pair correlation for the zero mode $\lvert\mathbf{k}\rvert=0$ over the first three inflaton oscillations for the collisionless case, i.e.~with $\lambda = 0$. We see that the density matrix approach correctly captures the resonant particle production, and it therefore provides a framework within which to study non-adiabatic particle production that is complementary to existing methods based on solving the Mathieu equation for the field modes. In particular, we see the characteristic jumps in the number density, occurring each time the inflaton field passes through zero. Note that the adiabatic approximation is satisfied between each jump, i.e.~$\lvert\dot{\omega}_{\mathbf{k}}/\omega^{2}_{\mathbf{k}}\rvert<1$, such that $N_{\mathbf{k}}$ is an approximate adiabatic invariant and remains roughly constant. From Eq.~\eqref{final eoms 1}, it is clear that the pair correlations act to source the growth in the number density. This is corroborated by the numerics, where, in Fig.~\ref{fig:free n M evo}, we see that the growth in the pair correlations precedes the growth in the number density. We reiterate that the presence of the pair correlations plays a crucial role in the non-adiabatic particle production.

\begin{figure}[htbp!]
\begin{center}

\includegraphics[width=0.75\textwidth]{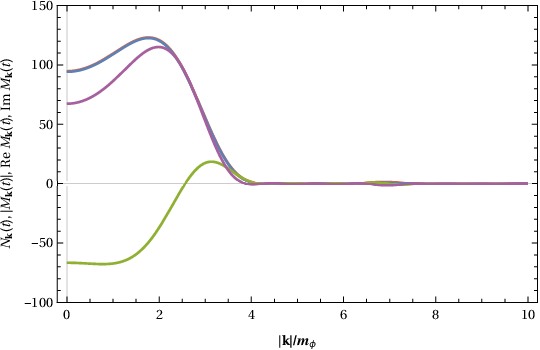}

\end{center}

\caption{Plots of $N_{\mathbf{k}}$ (blue), $\lvert M_{\mathbf{k}}\rvert$ (orange), $\text{Re}\,M_{\mathbf{k}}$ (green) and $\text{Im}\,M_{\mathbf{k}}$ (purple) as a function of $\lvert\mathbf{k}\rvert /m_{\phi}$ at $t = 4\pi/m_{\phi}$.}\label{fig:free n M spectrum}

\end{figure}

Figure~\ref{fig:free n M spectrum} shows plots of $N_{\mathbf{k}}$, $\lvert M_{\mathbf{k}}\rvert$, $\text{Re}\,M_{\mathbf{k}}$ and $\text{Im}\,M_{\mathbf{k}}$ as a function of $\lvert\mathbf{k}\rvert /m_{\phi}$ at $t = 4\pi/m_{\phi}$. We see that the number density is non-zero for a continuous range of momenta, typically within the interval $0\leq\lvert \mathbf{k}\rvert\lesssim m_{\phi}(q/4)^{1/4}\approx\;4m_{\phi}$, as is expected for broad resonance. Importantly,  Fig.~\ref{fig:free n M spectrum} also confirms the expected result that $N_{\mathbf{k}}$ and $\lvert M_{\mathbf{k}}\rvert$ are the same order of magnitude throughout the preheating phase. In fact, they are almost identical.  


\begin{figure}[t!]

\vspace{2em}

\begin{center}
  \subfloat[\label{fig:deltaN_k_lambda_01_02}]{\includegraphics[width=0.75\textwidth]{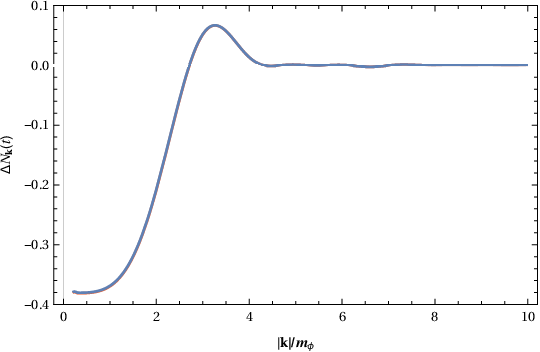}}\vfill\vfill
  \subfloat[\label{fig:N_evo_lambda_0_01_02}]{\includegraphics[width=0.75\textwidth]{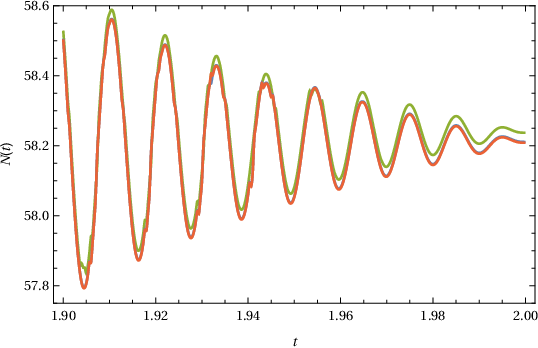}}
\end{center}

\caption{(a) Difference $\Delta N_{\mathbf{k}}=N^{\lambda}_{\mathbf{k}}-N^{\lambda\, =\, 0}_{\mathbf{k}}$ between the collisional and collisionless number densities as a function of $|\mathbf{k}|/m_{\phi}$ for $\lambda=0.1$ (blue) and $\lambda=0.2$ (orange) at $t=4\pi/m_{\phi}$ and in the case where only the number density $N_{\mathbf{k}}$ participates in the collision integral. (b) Time evolution of the total number per unit volume~$N(t)$ for $\lambda=0$ (green), $\lambda=0.1$ (blue) and $\lambda=0.2$ (orange), where the time evolution is shown in units of $m_{\phi}/2\pi$ (corresponding to the number of periods of the inflaton condensate). We have truncated the graph near to $t\sim 4\pi/m_{\phi}$ to make the suppression visible.}

\end{figure}

We now turn our attention to the collisional cases, i.e.~$\lambda\neq 0$. In the first instance, we set the pair correlations $M$ and $M^*$ to zero in the collision terms in Eq.~\eqref{final eoms 1}, so as to be able to isolate their impact. Figure~\ref{fig:deltaN_k_lambda_01_02} shows the number density as a function of $|\mathbf{k}|/ m_{\phi}$ for the collisionless case and collisional cases with $\lambda \in\{ 0.1, 0.2\}$, neglecting the pair correlations. While the maximum difference is at the sub-percent level ($\sim 0.3\%$) for both the $\lambda=0.1$ and $\lambda=0.2$ cases, we see that the collisions lead to a suppression of the particle production, corresponding to a reduction in the efficiency of the resonance, as we might expect. This suppression is also visible in Fig.~\ref{fig:N_evo_lambda_0_01_02}, where we show the time-evolution of the collisionless \\ \\ \\ and collisional number densities for the same case. These results illustrate that the collisions have an effect (albeit initially small) fairly soon after the onset of preheating. Importantly, one would expect these effects to become more pronounced as preheating proceeds and as the number density grows.

\begin{figure}[t!]

\vspace{2em}

\begin{center}
  \includegraphics[width=0.75\textwidth]{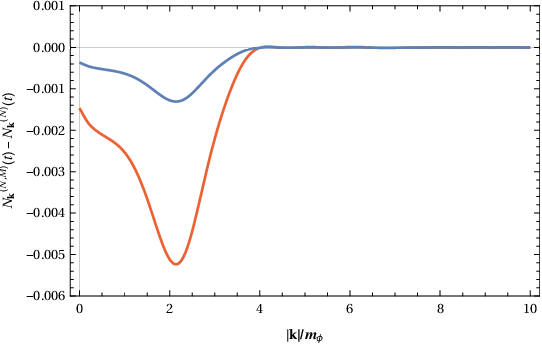}
\end{center}

\caption{Difference \smash{$N^{(N,M)}_{\mathbf{k}}-N^{(N)}_{\mathbf{k}}$} between the number densities with both $N_{\mathbf{k}}$ and $M_{\mathbf{k}}$ participating (\smash{$N^{(N,M)}_{\mathbf{k}}$}) in the collision terms, and with only $N_{\mathbf{k}}$ participating (\smash{$N^{(N)}_{\mathbf{k}}$}) for $\lambda=0.1$ (blue) and $\lambda=0.2$ (orange) at $t=4\pi/m_{\phi}$.}\label{fig:nkonlyn_nknandM_lambda_01_02}

\end{figure}

\begin{figure}[t!]

\vspace{2em}

\begin{center}

  \includegraphics[width=0.75\textwidth]{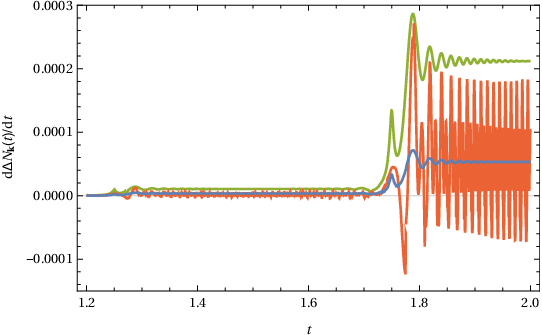}

\end{center}

\caption{Behaviour of the collision integral $\sd{d}\Delta N(t)/\sd{d}t$ as a function of time $t$ (in units of $m_{\phi}/2\pi$). The blue and green curves correspond to the case where only $N_{\mathbf{k}}$ contributes, for $\lambda \, =\, 0.1$ and $\lambda \, =\, 0.2$ respectively, and the orange curve, to the case in which $M_{\mathbf{k}}$ and $M^{\ast}_{\mathbf{k}}$ are also included. The graph has been truncated at $t\, =\, 1.2$, since the collision integral is negligible beforehand. In particular, it is found that the magnitude of $\sd{d}\Delta N(t)/\sd{d}t$ for $\lambda \, =\, 0.2$ is four times larger than the $\lambda \, =\, 0.1$, as would be expected.}\label{fig:deltandot}

\end{figure}

In Fig.~\ref{fig:nkonlyn_nknandM_lambda_01_02}, we plot the difference between the cases with and without the pair correlations for the collisional case with $\lambda=0.1$. The impact of the pair correlations is negligible, despite the magnitudes of $N_{\mathbf{k}}$ and $\lvert M_{\mathbf{k}}\rvert$ being almost identical. However, by plotting the time-evolution of the integrated collision term
\begin{equation}
 \frac{\sd{d}}{\sd{d}t}\,\Delta N(t)\ =\ \int\!\frac{\sd{d}^{3}\mathbf{k}}{(2\pi)^{3}}\;{\rm Re}\,\mathcal{C}^{(N)}_{\mathbf{k}}[N,M;t]
\end{equation}
for the cases with and without the pair correlations (see Fig.~\ref{fig:deltandot}), the reason for this negligible impact becomes apparent.  Specifically, the contributions from $M_{\mathbf{k}}$ and $M^{\ast}_{\mathbf{k}}$ result in highly oscillatory contributions, which fluctuate about an average value that is only negligibly different from that of the case where only $N_{\mathbf{k}}$ contributes. This can be traced back to the master equations for $M_{\mathbf{k}}$ and $M^{\ast}_{\mathbf{k}}$, obtained from Eqs.~\eqref{M eom} and~\eqref{Mstar eom}. Both contain an oscillatory contribution with instantaneous period $T\sim2\pi/\omega_{\mathbf{k}}(t)$. As such, $M_{\mathbf{k}}$ and $M^{\ast}_{\mathbf{k}}$ oscillate on time-scales much shorter than those over which collisional processes take place, and their contribution effectively averages to zero. 


Therefore, with the present separation of scales, we can safely ignore any contributions from $M_{\mathbf{k}}$ and $M^{\ast}_{\mathbf{k}}$ to the collision integrals in the master equations. On the other hand, if the collision rate becomes comparable to the rate of oscillation of the pair correlations, one might expect a greater residual effect on the evolution of the number density. However, one might then doubt the applicability of the approximations used here to treat the time-dependence of the phase space, and we leave dedicated studies to future work.

\begin{figure}[t!]

\vspace{3em}

\begin{center}

  \subfloat[\label{fig:nkfree6pi}]{\includegraphics[width=0.75\textwidth]{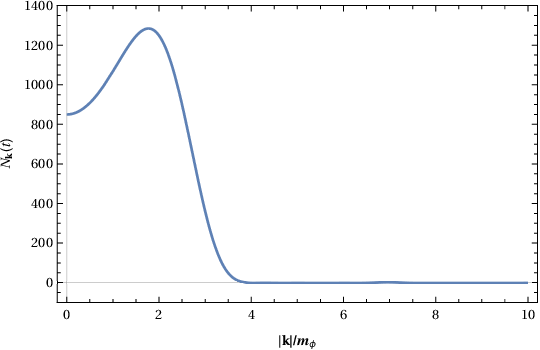}}\\ 

  \hspace{0.3cm}\subfloat[\label{fig:deltaN_k_lambda_02_2pi_4pi_6pi}]{\includegraphics[width=0.73\textwidth]{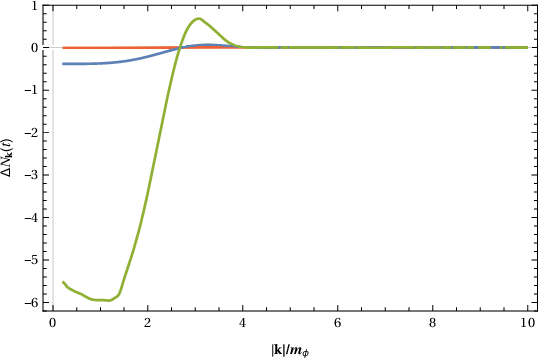}}
\end{center}

\caption{(a) Plot of the number density in the collisionless case $N^{\lambda\,=\,0}_{\mathbf{k}}$ at $t= 0.9\times 6\pi/m_{\phi}$. (b) The difference $\Delta N_{\mathbf{k}}=N^{\lambda}_{\mathbf{k}}-N^{\lambda\,=\,0}_{\mathbf{k}}$ between the collisional and collisionless number density for $\lambda=0.2$ at $t =  2\pi/m_{\phi}$ (orange), $t= 4\pi/m_{\phi}$ (blue) and $t =0.9\times 6\pi/m_{\phi}$ (green) in the case where only the number density $N_{\mathbf{k}}$ participates in the collision integral.\vspace{4em}}\label{fig:lambda 01}

\end{figure}

By neglecting the contributions from $M_{\mathbf{k}}$ and $M^{\ast}_{\mathbf{k}}$ to the collision integrals, as we have shown to be appropriate for the present choice of parameters, the stability of the numerics is improved. By this means, we were able to evolve the system reliably for three full oscillations of the inflaton field. The collisionless number density is shown in Fig.~\ref{fig:nkfree6pi}, having increased in amplitude by an order of magnitude compared with the previous inflaton oscillation shown in Fig.~\ref{fig:free n M spectrum}. The comparison with the collisional case is presented in Fig.~\ref{fig:deltaN_k_lambda_02_2pi_4pi_6pi}. We show how the suppression of the number density increases with each full oscillation of the inflaton field over the three periods. In particular, we see that after three periods, the maximum relative difference has increased from the sub-percent level ($\sim 0.3\%$) to order $1\%$, that is the effect of the collision integrals has essentially doubled after only one additional cycle.

\section{Conclusions}
\label{sec:conc}

Within the density matrix formalism, we have derived a (self-consistent) set of quantum Boltzmann equations, which are able to describe the evolution of an ensemble of self-inter\-acting scalar particles that are subject to an oscillating mass term. During the preheating phase of the early universe, these equations can be used to determine the evolution of the number density and pair correlations for a scalar field coupled to the inflaton, while accounting also for collisional processes. For a toy model of preheating, we have been able to solve for this evolution over the first few inflaton oscillations, and our conclusions can be summarized as follows:
\begin{itemize}

\item We have illustrated the importance of the pair correlations during preheating. In particular, we have shown that they play a crucial role in mediating the non-adiabatic particle production. Moreover, we have demonstrated that the pair correlations are of comparable magnitude to the number density throughout preheating and cannot, therefore, be neglected \emph{a priori} in the collision integrals that precipitate thermalization. 

\item In spite of the latter observation, we have, however, also demonstrated that the pair correlations can safely be neglected in the collision integrals when the collision rate is much slower than the rate of free-phase oscillations of the pair correlations. When this is the case, the contributions of the pair correlations are effectively time-averaged away.

\item Perhaps most importantly, our numerical analysis suggests that the collision terms have an impact on the resulting number density even in the very early stages of preheating. Specifically, after only three oscillations of the inflaton condensate, we find an $\mathcal{O}(1\%)$ deviation in the magnitude of the number density compared to the collisionless case. This deviation is expected to increase significantly as preheating progresses, and this motivates further numerical studies beyond the present work to establish the effect of accounting fully for the thermalization processes during preheating (and reheating) on the thermal history of the early universe.

\end{itemize}

Here, we have considered a simplified toy model of preheating. In a more realistic scenario, one would need to account for the effects of the Hubble expansion both on the decay of the inflaton condensate and the structure of the resonance bands. In addition, one would want to account for perturbative inflaton decays, as well as the backreaction of the particle-production processes on the inflaton condensate. In the case of broad resonance, the cosmological expansion results in a stochastic resonance behaviour, wherein the number density only increases exponentially on average~\cite{Kofman:1997yn,Mukhanov:2005sc}. This background expansion competes with the effects of backreactions and rescatterings in determining the efficiency of the resonant particle production~\cite{Mukhanov:2005sc}. One may therefore anticipate the effects of collisional processes during the production phases to remain significant in more realistic scenarios. We leave the (technically challenging) exploration of these possibilities for future research. Finally, we remark that it would be constructive to make direct comparisons between the present density matrix approach, where the relation to canonical quantities such as the number density is manifest, and others based on the closed-time-path formalism of non-equilibrium quantum field theory, where one must instead employ, e.g., quasi-particle approximations in order to extract physical observables.


\acknowledgments

PM and PS would like to thank Ed Copeland for earlier and enjoyable collaboration on parallel aspects of the problem of preheating. PM would like to thank Anupam Mazumdar for helpful discussions. WE would like to thank Felipe Maykot for helpful discussions on the intricacies of Mathematica. This work was supported by the Science and Technology Facilities Council (STFC) under Grant No.~ST/L000393/1 and a Leverhulme Trust Research Leadership Award.


\appendix


\section{Derivation of the approximate mode functions}
\label{sec:appendix b}

In this appendix, we derive the approximate solution for the function $\alpha_{\mathbf{k}}(t',t)$, appearing in Subsec.~\ref{ssec:collisions}, that is needed for the evaluation of the collision integrals in Appendix~\ref{sec:appendix c}.

The Bogoliubov coefficients satisfy the first-order differential equations
\begin{subequations}
\begin{flalign}
\dot{\alpha}_{\mathbf{k}}(t,t')\ &=\ -\:i\omega_{\mathbf{k}}(t)\alpha_{\mathbf{k}}(t,t')\:+\:\frac{1}{2}\,\frac{\dot{\omega}_{\mathbf{k}}(t)}{\omega_{\mathbf{k}}(t)}\,\beta_{\mathbf{k}}(t,t')\;,\\ \dot{\beta}_{\mathbf{k}}(t,t')\ &=\ +\:i\omega_{\mathbf{k}}(t)\beta_{\mathbf{k}}(t,t')\:+\:\frac{1}{2}\,\frac{\dot{\omega}_{\mathbf{k}(t)}}{\omega_{\mathbf{k}}(t)}\,\alpha_{\mathbf{k}}(t,t')\;,
\end{flalign}
\end{subequations}
which follow from Eq.~\eqref{eq:aadagfull}. For $t\sim t'$, we can set $\beta_{\mathbf{k}}(t,t')\simeq 0$, such that
\begin{equation}
\label{eq:alphaapprox}
\dot{\alpha}_{\mathbf{k}}(t,t')\ \simeq\ -\:i\omega_{\mathbf{k}}(t)\alpha_{\mathbf{k}}(t,t')\;,
\end{equation}
effectively imposing adiabatic evolution in the neighbourhood of the time $t$. The solution to Eq.~\eqref{eq:alphaapprox}, satisfying the boundary condition in Eq.~\eqref{aa* initial conditions}, is
\begin{equation}\label{ab approx solutions 2}
 \alpha_{\mathbf{k}}(t,t')\ \simeq\ \text{exp}\bigg[-i\!\int_{t'}^{t}\mathrm{d}\tilde{t}\;\omega_{\mathbf{k}}(\tilde{t})\bigg]\;,
\end{equation}
and the inverse function $\alpha_{\mathbf{k}}(t',t)$ can be obtained from the property $\alpha_{\mathbf{k}}(t',t)=\alpha^{\ast}_{\mathbf{k}}(t,t')$.

\begin{figure}[t!]

\begin{center}

\includegraphics[width=0.75\textwidth]{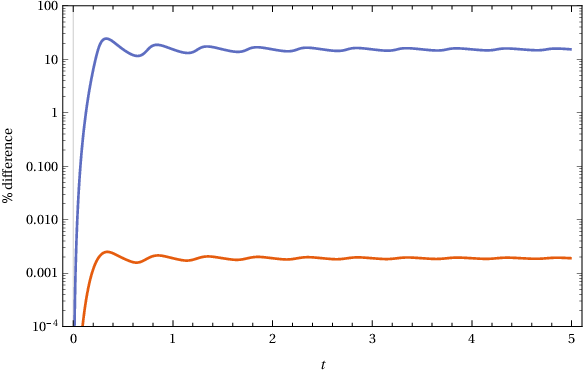}

\end{center} 

\caption[.]{Plot of the percentage difference between \mbox{$\int_{0}^{t}\sd{d}t'\,\omega_{\mathbf{k}}(t')$} and its Taylor approximation (cf.~Eq.~\eqref{approx elliptic}) as a function of $t$ (in units of $m_{\phi}/2\pi$) over several periods of oscillation of the inflaton condensate for \mbox{$z_{\mathbf{k}}\sim\mathcal{O}(1)$} (blue line) and \mbox{$z_{\mathbf{k}}\ll 1$} (orange).\label{fig:elliptic12}}

\end{figure}

The integrand in the exponent of Eq.~\eqref{ab approx solutions 2} can be evaluated in closed form; specifically,
\begin{equation}\label{elliptic solution}
\int_{t'}^{t}\sd{d}\tilde{t}\;\omega_{\mathbf{k}}(\tilde{t})\ =\ \frac{\omega_{\mathbf{k}}\vert_{\text{max}}}{m_{\phi}}\left[\text{E}\left(m_{\phi}t,z_{\mathbf{k}}\right)\:-\:\text{E}\left(m_{\phi}t',z_{\mathbf{k}}\right)\right]\;,
\end{equation}
where $\text{E}\left(m_{\phi}t,z_{\mathbf{k}}\right)$ is the incomplete elliptic integral of the second kind, $z_{\mathbf{k}}=g\varphi_{0}^{2}/[2(\omega_{\mathbf{k}}\vert_{\text{max}})^{2}]$ and
\begin{equation}
\omega_{\mathbf{k}}\vert_{\text{max}}\ \coloneqq\ \sqrt{\mathbf{k}^{2}+m_{\chi}^{2}+\frac{g\varphi_{0}^{2}}{2}}
\end{equation}
is the maximum value of $\omega_{\mathbf{k}}(t)$. While it is not appropriate to expand $\omega_{\mathbf{k}}(t)$ perturbatively in the coupling $g$, since $g\varphi_0^2\gg m_{\chi}^2$, a numerical analysis of Eq.~\eqref{elliptic solution} suggests that it is possible to approximate the solution well by making a series expansion of $\text{E}\left(m_{\phi}t,z_{\mathbf{k}}\right)$ with respect to $z_{\mathbf{k}}$ (for $z_{\mathbf{k}}<1$). To linear order, we obtain
\begin{equation}\label{approx elliptic}
 \text{E}\left(m_{\phi}t,z_{\mathbf{k}}\right)\:-\:\text{E}\left(m_{\phi}t',z_{\mathbf{k}}\right)\ \simeq\ \left(1-\frac{z_{\mathbf{k}}}{4}\right)m_{\phi}(t-t')\:+\:\frac{z_{\mathbf{k}}}{4}m_{\phi}\big[t\,\text{sinc}(2m_{\phi}t)\:-\:t'\,\text{sinc}(2m_{\phi}t')\big]\;.
\end{equation}
For $z_{\mathbf{k}}\ll 1$, the relative error is $\sim 0.01\%$; for $z_{\mathbf{k}}\sim\mathcal{O}(1)$, it is at most $\sim 15\%$ (see Fig.~\ref{fig:elliptic12}). Away from $t,t'=0$, the first term dominates, and we can neglect the terms involving ${\rm sinc}$ functions, whose relative contributions decrease linearly with time for $t>1/m_{\phi}$. Doing so, we are left with the result
\begin{equation}\label{elliptic solution approx}
 \int_{t'}^{t}\mathrm{d}\tilde{t}\;\omega_{\mathbf{k}}(\tilde{t})\ \simeq\ \bar{\omega}_{\mathbf{k}}(t-t')\;,
\end{equation}
where $\bar{\omega}_{\mathbf{k}}$, as given by Eq.~\eqref{time-average}, is the (approximate) time-average of $\omega_{\mathbf{k}}(t)$ over an interval $\Delta t= t-t'$, i.e.
\begin{equation}
\bar{\omega}_{\mathbf{k}}\ \simeq\ \frac{1}{\Delta t}\int_{t'}^{t}\mathrm{d}\tilde{t}\;\omega_{\mathbf{k}}(\tilde{t})\;.
\end{equation}
To be consistent with this approximation, we should make the replacement $\omega_{\mathbf{k}}(t)\,\to\,\bar{\omega}_{\mathbf{k}}$ in the mode functions (cf.~Eq.~\eqref{mode functions}) and whenever it appears in the collision integrals. In this way, Eq.~\eqref{mode functions} for the mode function reduces to
\begin{equation}\label{a+b approx elliptic}
 \tilde{\chi}_{\mathbf{k}}(t',t)\ \simeq\ \frac{1}{\sqrt{2\bar{\omega}_{\mathbf{k}}}}\,e^{+i\bar{\omega}_{\mathbf{k}}(t-t')}\;.
\end{equation}
Notice that, in the limit $\varphi_0\to 0$, corresponding to the decay of the amplitude of the inflaton oscillations, $\bar{\omega}_{\mathbf{k}}\to \sqrt{\mathbf{k}^2+m_{\chi}^2}$, and we correctly recover the usual evolution of the free field.

\section{Derivation of the collision terms}
\label{sec:appendix c}

In this appendix, we present further technical details of the derivation of the collision terms appearing in the Boltzmann equations in Subsec.~\ref{ssec:master eqs}.

The various collision terms are given by the expression
\begin{align}
\mathcal{C}^{(\mathcal{O})}_{\mathbf{k}}[N,M;t]\ &=\ -\:\frac{1}{2}\int_{-\infty}^{+\infty}{\rm d}t'\;\big\langle\big[\big[\hat{\mathcal{O}}_{\mathbf{k}}(t),\hat{H}_{\text{int}}(t)\big],\hat{H}_{\text{int}}(t')\big]\big\rangle_{t}\nonumber\\ &=\ -\:\frac{1}{2}\bigg(\frac{\lambda}{4!}\bigg)^{\!2}\int_{-\infty}^{+\infty}{\rm d}t'\int\mathrm{d}^{3}\mathbf{x}\,\mathrm{d}^{3}\mathbf{y}\,\big\langle\big[\big[\hat{\mathcal{O}}_{\mathbf{k}}(t),\hat{\chi}^{4}(x)\big],\hat{\chi}^{4}(y)\big]\big\rangle_{t}\;,
\end{align}
where $\hat{\mathcal{O}}_{\mathbf{k}}\in\{\hat{N}_{\mathbf{k}},\hat{M}_{\mathbf{k}}\}$, and $x^{\mu}=(t,\mathbf{x})$ and $y^{\mu}=(t',\mathbf{y})$.  The double commutator can be expanded as follows:
\begin{flalign}
 \big[\big[\hat{\mathcal{O}}_{\mathbf{k}}(t),\,\hat{\chi}^{4}(t,\mathbf{x})\big],\,\hat{\chi}^{4}(t',\mathbf{y})\big]\ &=\ 4\big[\big[\hat{\mathcal{O}}_{\mathbf{k}},\,\hat{\chi}\big]\hat{\chi}^{3},\,\hat{\chi}'^{4}\big]\:-\:6\Delta_{\hat{\mathcal{O}}_{\mathbf{k}}}\big[\hat{\chi}^{2},\,\hat{\chi}'^{4}\big]\qquad\qquad\qquad\nonumber\\ \ &=\ 4\big[\hat{\mathcal{O}}_{\mathbf{k}},\,\hat{\chi}\big]\big[\hat{\chi}^{3},\,\hat{\chi}'^{4}\big]\:+\:4\big[\big[\hat{\mathcal{O}}_{\mathbf{k}},\,\hat{\chi}\big],\,\hat{\chi}'^{4}\big]\hat{\chi}^{3}\: -\:6\Delta_{\hat{\mathcal{O}}_{\mathbf{k}}}\big[\hat{\chi}^{2},\,\hat{\chi}'^{4}\big]\nonumber\\ \ &=\ 16\Delta'_{\hat{\mathcal{O}}_{\mathbf{k}}}(\hat{\chi}^{3}\hat{\chi}'^{3}\:-\:9\Delta\hat{\chi}^{2}\hat{\chi}'^{2}\:+\:18\Delta^{2}\hat{\chi}\hat{\chi}'-6\Delta^{3}\big)\nonumber\\&\qquad +\:48\Delta\big[\hat{\mathcal{O}}_{\mathbf{k}},\,\hat{\chi}\big]\hat{\chi}^{2}\hat{\chi}'^{3}\:-\:144\Delta^{2}\big[\hat{\mathcal{O}}_{\mathbf{k}},\,\hat{\chi}\big]\hat{\chi}\hat{\chi}'^{2}\nonumber\\ &\qquad+\:96\Delta^{3}\big[\hat{\mathcal{O}}_{\mathbf{k}},\,\hat{\chi}\big]\hat{\chi}'\:-\:24\Delta_{\hat{\mathcal{O}}_{\mathbf{k}}}\Delta\big(2\hat{\chi}\hat{\chi}'^{3}\:-\:3\Delta\hat{\chi}'^{2}\big)\;,
\end{flalign}
where all operators with the latest time $t>t'$ have been commuted to the left (to simplify their Wick contraction). Herein, we have suppressed the spacetime arguments of the various operators and used the shorthand notations $\hat{\chi}\equiv \hat{\chi}(t,\mathbf{x})$ and $\hat{\chi}'\equiv \hat{\chi}(t',\mathbf{y})$. In addition, we have defined the various Pauli-Jordan-like functions
\begin{subequations}
\begin{flalign}
 \Delta\ &\equiv\ \Delta(x,y)\ \coloneqq\ \big[\hat{\chi}(x),\,\hat{\chi}(y)\big]\;,\\ 
 \Delta'_{\hat{\mathcal{O}}_{\mathbf{k}}}\ &\equiv\ \Delta_{\hat{\mathcal{O}}_{\mathbf{k}}}(x,y)\ \coloneqq\ \big[\big[\hat{\mathcal{O}}_{\mathbf{k}}(t),\,\hat{\chi}(x)\big],\,\hat{\chi}(y)\big]\;,\\ 
 \Delta_{\hat{\mathcal{O}}_{\mathbf{k}}}\ &\equiv\ \Delta_{\hat{\mathcal{O}}_{\mathbf{k}}}(x,x)\;.
\end{flalign}
\end{subequations}
After taking the trace with the density operator and performing the  Wick contractions, we obtain
\begin{flalign}
\label{double comm correlator}
 \mathcal{C}^{(\mathcal{O})}_{\mathbf{k}}[N,M;t]\ &=\ \frac{1}{2}\,\int_{-\infty}^{+\infty}{\rm d}t'\int_{\mathbf{x},\mathbf{y}}\bigg(\Big[\Pi^{<}(x,y)G^{>}_{\mathcal{O}_\mathbf{k}}(y,x)\:-\:\Pi^{>}(x,y)G^{<}_{\mathcal{O}_\mathbf{k}}(y,x)\Big]\nonumber\\ &\qquad+\:\frac{\lambda^{4}}{2}\Big[G^{<}_{\mathcal{O}_\mathbf{k}}(y,x)G^{>}(x,y)\:-\:G^{>}_{\mathcal{O}_\mathbf{k}}(y,x)G^{<}(x,y)\Big]\big\langle\hat{\chi}^{2}(x)\big\rangle_{t}\big\langle\hat{\chi}^{2}(y)\big\rangle_{t}\nonumber\\ &\qquad+\:\frac{\lambda^{2}}{8}\,\Big[G^{<}_{\mathcal{O}_\mathbf{k}}(x,x)\big(G^{>}(x,y)\big)^{2}\:-\:G^{<}_{\mathcal{O}_\mathbf{k}}(x,x)\big(G^{<}(x,y)\big)^{2}\nonumber\\ &\qquad\quad\;\;\;+\:G^{>}_{\mathcal{O}_\mathbf{k}}(x,x)\big(G^{<}(x,y)\big)^{2}\:-\:G^{>}_{\mathcal{O}_\mathbf{k}}(x,x)\big(G^{>}(x,y)\big)^{2}\Big]\big\langle\hat{\chi}^{2}(y)\big\rangle_{t}\bigg)\;,
\end{flalign}
where 
\begin{subequations}
\begin{align}
&G^{>}(x,y)\ \coloneqq\ \langle\chi(x)\chi(y)\rangle_{t}\nonumber\\ &\qquad =\ \int_{\mathbf{p}}\Big[\tilde{\chi}_{\mathbf{p}}(t',t)e^{+i\mathbf{p}\cdot\left(\mathbf{x}-\mathbf{y}\right)}\big(N_{\mathbf{p}}(t)+M_{\mathbf{p}}(t)\big)+\tilde{\chi}^{\ast}_{\mathbf{p}}(t',t)e^{-i\mathbf{p}\cdot\left(\mathbf{x}-\mathbf{y}\right)}\big(1+N_{\mathbf{p}}(t)+M^{\ast}_{\mathbf{p}}(t)\big)\Big]\;,\\
&G^{<}(x,y)\ \coloneqq\ \langle\chi(y)\chi(x)\rangle_{t}\nonumber\\ &\qquad =\ \int_{\mathbf{p}}\Big[\tilde{\chi}_{\mathbf{p}}(t',t)e^{+i\mathbf{p}\cdot\left(\mathbf{x}-\mathbf{y}\right)}\big(1+N_{\mathbf{p}}(t)+M_{\mathbf{p}}(t)\big)+\tilde{\chi}^{\ast}_{\mathbf{p}}(t',t)e^{-i\mathbf{p}\cdot\left(\mathbf{x}-\mathbf{y}\right)}\big(N_{\mathbf{p}}(t)+M^{\ast}_{\mathbf{p}}(t)\big)\Big]
\end{align}
\end{subequations}
are the positive- and negative-frequency Wightman propagators, and $\Pi^{\lessgtr}(x,y)$ and $G_{\mathcal{O}_{\mathbf{k}}}^{\lessgtr}(x,y)$ are defined in Eq.~\eqref{wightman}. For illustration, we have
\begin{subequations}
 \begin{flalign}
  G^{>}_{N_\mathbf{k}}(y,x)\ &=\  \frac{1}{\text{Vol}\sqrt{2\omega_{\mathbf{k}}(t)}}\,\bigg(\Big[\tilde{\chi}_{\mathbf{k}}(t',t)\big(1+N_{\mathbf{k}}(t)\big)+\tilde{\chi}^{\ast}_{\mathbf{k}}(t',t)M^{\ast}_{\mathbf{k}}(t)\Big]e^{-i\mathbf{k}\cdot\left(\mathbf{x}-\mathbf{y}\right)}\nonumber\\ &\qquad\qquad\qquad\quad -\Big[\tilde{\chi}_{\mathbf{k}}(t',t)M_{\mathbf{k}}(t)+\tilde{\chi}^{\ast}_{\mathbf{k}}(t',t)N_{\mathbf{k}}(t)\Big]e^{+i\mathbf{k}\cdot\left(\mathbf{x}-\mathbf{y}\right)}\bigg)\nonumber\\&=\ -\:\Big[G^{<}_{N_\mathbf{k}}(y,x)\Big]^{\ast} ,\\[0.5em]
  G^{>}_{M_\mathbf{k}}(y,x)\ &=\  \frac{1}{\text{Vol}\sqrt{2\omega_{\mathbf{k}}(t)}}\,\Big[\tilde{\chi}_{\mathbf{k}}(t',t)M_{\mathbf{k}}(t)+\tilde{\chi}^{\ast}_{\mathbf{k}}(t',t)N_{\mathbf{k}}(t)\Big]\Big(e^{+i\mathbf{k}\cdot\left(\mathbf{x}-\mathbf{y}\right)}+e^{-i\mathbf{k}\cdot\left(\mathbf{x}-\mathbf{y}\right)}\Big)\,,\\[0.5em] G^{<}_{M_\mathbf{k}}(y,x)\ &=\  \frac{1}{\text{Vol}\sqrt{2\omega_{\mathbf{k}}(t)}}\,\Big[\tilde{\chi}_{\mathbf{k}}(t',t)M_{\mathbf{k}}(t)+\tilde{\chi}^{\ast}_{\mathbf{k}}(t',t)(1+N_{\mathbf{k}}(t))\Big]\Big(e^{+i\mathbf{k}\cdot\left(\mathbf{x}-\mathbf{y}\right)}+e^{-i\mathbf{k}\cdot\left(\mathbf{x}-\mathbf{y}\right)}\Big)\, .
 \end{flalign}
\end{subequations}
The collision terms arise from the first line of Eq.~\eqref{double comm correlator} and, in the Markovian limit, these terms give rise to the relevant two-to-two scattering processes. The remaining terms in Eq.~\eqref{double comm correlator} correspond to $\mathcal{O}(\lambda^{2})$ shifts in the mass of the $\chi$ field, which we omit from the present analysis.

Expanding the terms in the first line of Eq.~\eqref{double comm correlator} and inserting the approximate solution for $\alpha_{\mathbf{k}}(t',t)$ (cf.~Appendix~\ref{sec:appendix b}), we arrive at the following general expression for the collision integral:
\begin{equation}\label{collision terms swapped momenta}
\mathcal{C}_{\mathbf{k}}^{(\mathcal{O})}[N,M;t]\ \simeq \ \frac{\lambda^{2}}{2}\,\sum_{j\,=\,1}^{4}\int\!\sd{d}\Pi^{(j)}_{\mathbf{p},\mathbf{q},\mathbf{k}}\;f^{(\mathcal{O})}_{(j);\mathbf{p},\mathbf{q},\mathbf{k}}[N,M;t]\;,
\end{equation} 
where we have introduced the modified phase-space measure
\begin{equation}
 \sd{d}\Pi^{(j)}_{\mathbf{p},\mathbf{q},\mathbf{k}}\ =\ \frac{\sd{d}^{3}\mathbf{p}}{(2\pi)^{3}}\frac{\sd{d}^{3}\mathbf{q}}{(2\pi)^{3}}\,2\pi\,\delta\left(\Delta\omega_{j}\right)\,\prod_{\boldsymbol\kappa}\frac{1}{2\bar{\omega}_{\boldsymbol\kappa}}\;,
\end{equation}
with $\boldsymbol\kappa\in\lbrace\mathbf{k},\,\mathbf{p},\,\mathbf{q},\,(\mathbf{p}+\mathbf{q}-\mathbf{k})\rbrace$, and defined the set of functions $\lbrace \Delta\omega_{j} |j=1,2,3,4\rbrace$, with
\begin{subequations}\label{a functions}
\begin{flalign}
 \Delta\omega_{1}\ &=\ \bar{\omega}_{\mathbf{k}}\:+\:\bar{\omega}_{\mathbf{p}+\mathbf{q}-\mathbf{k}}\:-\:\bar{\omega}_{\mathbf{p}}\:-\:\bar{\omega}_{\mathbf{q}}\;,\\
 \Delta\omega_{2}\ &=\ \bar{\omega}_{\mathbf{k}}\:+\:\bar{\omega}_{-\mathbf{p}-\mathbf{q}-\mathbf{k}}\:+\:\bar{\omega}_{\mathbf{p}}\:+\:\bar{\omega}_{\mathbf{p}}\;,\\
 \Delta\omega_{3}\ &=\ \bar{\omega}_{\mathbf{p}}\:+\:\bar{\omega}_{\mathbf{q}}\:+\:\bar{\omega}_{\mathbf{k}-\mathbf{p}-\mathbf{q}}\:-\:\bar{\omega}_{\mathbf{k}}\;,\\
 \Delta\omega_{4}\ &=\ \bar{\omega}_{\mathbf{k}}\:+\:\bar{\omega}_{\mathbf{q}}\:+\:\bar{\omega}_{\mathbf{p}-\mathbf{k}-\mathbf{q}}\:-\:\bar{\omega}_{\mathbf{p}}\;.
\end{flalign}
\end{subequations}
Having made the Wigner-Weisskopf (or Markovian) approximation, we restore energy conservation at each interaction vertex. The only kinematically viable process is then the two-to-two scattering, corresponding to the case $j=1$ above.

The set of functions \smash{$\lbrace f^{(\mathcal{O})}_{(j);\mathbf{p},\mathbf{q},\mathbf{k}}[N,M;t]|j=1,2,3,4\rbrace$} contain the statistical factors, and their explicit expressions are as follows:
\begin{subequations}
\begin{flalign}
 f^{(N)}_{(1);\mathbf{p},\mathbf{q},\mathbf{k}}\ &=\ \left(1+N_{\mathbf{k}}\right)\left(N_{\mathbf{p}}+M^{\ast}_{\mathbf{p}}\right)\left(N_{\mathbf{q}}+M^{\ast}_{\mathbf{q}}\right)\left(1+N_{\mathbf{p}+\mathbf{q}-\mathbf{k}}+M_{\mathbf{p}+\mathbf{q}-\mathbf{k}}\right)\nonumber\\ &\qquad -\:N_{\mathbf{k}}\left(1+N_{\mathbf{p}}+M^{\ast}_{\mathbf{p}}\right)\left(1+N_{\mathbf{q}}+M^{\ast}_{\mathbf{q}}\right)\left(N_{\mathbf{p}+\mathbf{q}-\mathbf{k}}+M_{\mathbf{p}+\mathbf{q}-\mathbf{k}}\right)\nonumber\\ & \qquad +\:M^{\ast}_{\mathbf{k}}\left(1+N_{\mathbf{p}}+M_{\mathbf{p}}\right)\left(1+N_{\mathbf{q}}+M_{\mathbf{q}}\right)\left(N_{\mathbf{p}+\mathbf{q}-\mathbf{k}}+M^{\ast}_{\mathbf{p}+\mathbf{q}-\mathbf{k}}\right)\nonumber\\ &\qquad -\:M^{\ast}_{\mathbf{k}}\left(N_{\mathbf{p}}+M_{\mathbf{p}}\right)\left(N_{\mathbf{q}}+M_{\mathbf{q}}\right)\left(1+N_{\mathbf{p}+\mathbf{q}-\mathbf{k}}+M^{\ast}_{\mathbf{p}+\mathbf{q}-\mathbf{k}}\right)\;,\nonumber \\ \\
f^{(N)}_{(2);\mathbf{p},\mathbf{q},\mathbf{k}}\ &=\ \frac{1}{3}\Big[\left(1+N_{\mathbf{k}}\right)\left(1+N_{\mathbf{p}}+M_{\mathbf{p}}\right)\left(1+N_{\mathbf{q}}+M_{\mathbf{q}}\right)\left(1+N_{-\mathbf{k}-\mathbf{p}-\mathbf{q}}+M_{-\mathbf{k}-\mathbf{p}-\mathbf{q}}\right)\nonumber\\ & \qquad-\:N_{\mathbf{k}}\left(N_{\mathbf{p}}+M_{\mathbf{p}}\right)\left(N_{\mathbf{q}}+M_{\mathbf{q}}\right)\left(N_{-\mathbf{k}-\mathbf{p}-\mathbf{q}}+M_{-\mathbf{k}-\mathbf{p}-\mathbf{q}}\right)\nonumber\\ &\qquad +\:M^{\ast}_{\mathbf{k}}\left(N_{\mathbf{p}}+M^{\ast}_{\mathbf{p}}\right)\left(N_{\mathbf{q}}+M^{\ast}_{\mathbf{q}}\right)\left(N_{-\mathbf{k}-\mathbf{p}-\mathbf{q}}+M^{\ast}_{-\mathbf{k}-\mathbf{p}-\mathbf{q}}\right)\nonumber\\ &\qquad -\:M^{\ast}_{\mathbf{k}}\left(1+N_{\mathbf{p}}+M^{\ast}_{\mathbf{p}}\right)\left(1+N_{\mathbf{q}}+M^{\ast}_{\mathbf{q}}\right)\left(1+N_{-\mathbf{k}-\mathbf{p}-\mathbf{q}}+M^{\ast}_{-\mathbf{k}-\mathbf{p}-\mathbf{q}}\right)\Big]\;,\nonumber\\ \\
f^{(N)}_{(3);\mathbf{p},\mathbf{q},\mathbf{k}}\ &=\ \frac{1}{3}\Big[\left(1+N_{\mathbf{k}}\right)\left(N_{\mathbf{p}}+M^{\ast}_{\mathbf{p}}\right)\left(N_{\mathbf{q}}+M^{\ast}_{\mathbf{q}}\right)\left(N_{\mathbf{k}-\mathbf{p}-\mathbf{q}}+M^{\ast}_{\mathbf{k}-\mathbf{p}-\mathbf{q}}\right)\nonumber\\ &\qquad -\:N_{\mathbf{k}}\left(1+N_{\mathbf{p}}+M^{\ast}_{\mathbf{p}}\right)\left(1+N_{\mathbf{q}}+M^{\ast}_{\mathbf{q}}\right)\left(1+N_{\mathbf{k}-\mathbf{p}-\mathbf{q}}+M^{\ast}_{\mathbf{k}-\mathbf{p}-\mathbf{q}}\right)\nonumber\\ & \qquad+\:M^{\ast}_{\mathbf{k}}\left(1+N_{\mathbf{p}}+M_{\mathbf{p}}\right)\left(1+N_{\mathbf{q}}+M_{\mathbf{q}}\right)\left(1+N_{\mathbf{k}-\mathbf{p}-\mathbf{q}}+M_{\mathbf{k}-\mathbf{p}-\mathbf{q}}\right)\nonumber\\ & \qquad-\:M^{\ast}_{\mathbf{k}}\left(N_{\mathbf{p}}+M_{\mathbf{p}}\right)\left(N_{\mathbf{q}}+M_{\mathbf{q}}\right)\left(N_{\mathbf{k}-\mathbf{p}-\mathbf{q}}+M_{\mathbf{k}-\mathbf{p}-\mathbf{q}}\right)\Big]\;,\nonumber\\  \\
f^{(N)}_{(4);\mathbf{p},\mathbf{q},\mathbf{k}}\ &=\ \left(1+N_{\mathbf{k}}\right)\left(N_{\mathbf{p}}+M^{\ast}_{\mathbf{p}}\right)\left(1+N_{\mathbf{q}}+M_{\mathbf{q}}\right)\left(1+N_{\mathbf{p}-\mathbf{k}-\mathbf{q}}+M_{\mathbf{p}-\mathbf{k}-\mathbf{q}}\right)\nonumber\\ & \qquad-\:N_{\mathbf{k}}\left(1+N_{\mathbf{p}}+M^{\ast}_{\mathbf{p}}\right)\left(N_{\mathbf{q}}+M_{\mathbf{q}}\right)\left(N_{\mathbf{p}-\mathbf{k}-\mathbf{q}}+M_{\mathbf{p}-\mathbf{k}-\mathbf{q}}\right)\nonumber\\ &\qquad +\:M^{\ast}_{\mathbf{k}}\left(1+N_{\mathbf{p}}+M_{\mathbf{p}}\right)\left(N_{\mathbf{q}}+M^{\ast}_{\mathbf{q}}\right)\left(N_{\mathbf{p}-\mathbf{k}-\mathbf{q}}+M^{\ast}_{\mathbf{p}-\mathbf{k}-\mathbf{q}}\right)\nonumber\\ &\qquad -\:M^{\ast}_{\mathbf{k}}\left(N_{\mathbf{p}}+M_{\mathbf{p}}\right)\left(1+N_{\mathbf{q}}+M^{\ast}_{\mathbf{q}}\right)\left(1+N_{\mathbf{p}-\mathbf{k}-\mathbf{q}}+M^{\ast}_{\mathbf{p}-\mathbf{k}-\mathbf{q}}\right)\;, \nonumber\\  \\
 f^{(M)}_{(1);\mathbf{p},\mathbf{q},\mathbf{k}}\ &=\ N_{\mathbf{k}}\left(1+N_{\mathbf{p}}+M_{\mathbf{p}}\right)\left(1+N_{\mathbf{q}}+M_{\mathbf{q}}\right)\left(N_{\mathbf{p}+\mathbf{q}-\mathbf{k}}+M^{\ast}_{\mathbf{p}+\mathbf{q}-\mathbf{k}}\right)\nonumber\\ &\qquad -\:\left(1+N_{\mathbf{k}}\right)\left(N_{\mathbf{p}}+M_{\mathbf{p}}\right)\left(N_{\mathbf{q}}+M_{\mathbf{q}}\right)\left(1+N_{\mathbf{p}+\mathbf{q}-\mathbf{k}}+M^{\ast}_{\mathbf{p}+\mathbf{q}-\mathbf{k}}\right)\nonumber\\ &\qquad +\:M_{\mathbf{k}}\left(N_{\mathbf{p}}+M^{\ast}_{\mathbf{p}}\right)\left(N_{\mathbf{q}}+M^{\ast}_{\mathbf{q}}\right)\left(1+N_{\mathbf{p}+\mathbf{q}-\mathbf{k}}+M_{\mathbf{p}+\mathbf{q}-\mathbf{k}}\right)\nonumber\\ & \qquad -\:M_{\mathbf{k}}\left(1+N_{\mathbf{p}}+M^{\ast}_{\mathbf{p}}\right)\left(1+N_{\mathbf{q}}+M^{\ast}_{\mathbf{q}}\right)\left(N_{\mathbf{p}+\mathbf{q}-\mathbf{k}}+M_{\mathbf{p}+\mathbf{q}-\mathbf{k}}\right)\;,\nonumber\\  \\ 
f^{(M)}_{(2);\mathbf{p},\mathbf{q},\mathbf{k}}\ &=\ \frac{1}{3}\Big[N_{\mathbf{k}}\left(N_{\mathbf{p}}+M^{\ast}_{\mathbf{p}}\right)\left(N_{\mathbf{q}}+M^{\ast}_{\mathbf{q}}\right)\left(N_{-\mathbf{k}-\mathbf{p}-\mathbf{q}}+M^{\ast}_{-\mathbf{k}-\mathbf{p}-\mathbf{q}}\right)\nonumber\\ &\qquad -\:\left(1+N_{\mathbf{k}}\right)\left(1+N_{\mathbf{p}}+M^{\ast}_{\mathbf{p}}\right)\left(1+N_{\mathbf{q}}+M^{\ast}_{\mathbf{q}}\right)\left(1+N_{-\mathbf{k}-\mathbf{p}-\mathbf{q}}+M^{\ast}_{-\mathbf{k}-\mathbf{p}-\mathbf{q}}\right)\nonumber\\ &\qquad +\:M_{\mathbf{k}}\left(1+N_{\mathbf{p}}+M_{\mathbf{p}}\right)\left(1+N_{\mathbf{q}}+M_{\mathbf{q}}\right)\left(1+N_{-\mathbf{k}-\mathbf{p}-\mathbf{q}}+M_{-\mathbf{k}-\mathbf{p}-\mathbf{q}}\right)\nonumber\\ &\qquad -\:M_{\mathbf{k}}\left(N_{\mathbf{p}}+M_{\mathbf{p}}\right)\left(N_{\mathbf{q}}+M_{\mathbf{q}}\right)\left(N_{-\mathbf{k}-\mathbf{p}-\mathbf{q}}+M_{-\mathbf{k}-\mathbf{p}-\mathbf{q}}\right)\Big]\;,\nonumber\displaybreak \\  \\
f^{(M)}_{(3);\mathbf{p},\mathbf{q},\mathbf{k}}\ &=\ \frac{1}{3}\Big[N_{\mathbf{k}}\left(1+N_{\mathbf{p}}+M_{\mathbf{p}}\right)\left(1+N_{\mathbf{q}}+M_{\mathbf{q}}\right)\left(1+N_{\mathbf{k}-\mathbf{p}-\mathbf{q}}+M_{\mathbf{k}-\mathbf{p}-\mathbf{q}}\right)\nonumber\\ &\qquad -\:\left(1+N_{\mathbf{k}}\right)\left(N_{\mathbf{p}}+M_{\mathbf{p}}\right)\left(N_{\mathbf{q}}+M_{\mathbf{q}}\right)\left(N_{\mathbf{k}-\mathbf{p}-\mathbf{q}}+M_{\mathbf{k}-\mathbf{p}-\mathbf{q}}\right)\nonumber\\ &\qquad +\:M_{\mathbf{k}}\left(N_{\mathbf{p}}+M^{\ast}_{\mathbf{p}}\right)\left(N_{\mathbf{q}}+M^{\ast}_{\mathbf{q}}\right)\left(N_{\mathbf{k}-\mathbf{p}-\mathbf{q}}+M^{\ast}_{\mathbf{k}-\mathbf{p}-\mathbf{q}}\right)\nonumber\\ &\qquad -\:M_{\mathbf{k}}\left(1+N_{\mathbf{p}}+M^{\ast}_{\mathbf{p}}\right)\left(1+N_{\mathbf{q}}+M^{\ast}_{\mathbf{q}}\right)\left(1+N_{\mathbf{k}-\mathbf{p}-\mathbf{q}}+M^{\ast}_{\mathbf{k}-\mathbf{p}-\mathbf{q}}\right)\Big]\;, \nonumber\\ \\
f^{(M)}_{(4);\mathbf{p},\mathbf{q},\mathbf{k}}\ &=\ N_{\mathbf{k}}\left(1+N_{\mathbf{p}}+M_{\mathbf{p}}\right)\left(N_{\mathbf{q}}+M^{\ast}_{\mathbf{q}}\right)\left(N_{\mathbf{p}-\mathbf{k}-\mathbf{q}}+M^{\ast}_{\mathbf{p}-\mathbf{k}-\mathbf{q}}\right)\nonumber\\ &\qquad -\:\left(1+N_{\mathbf{k}}\right)\left(N_{\mathbf{p}}+M_{\mathbf{p}}\right)\left(1+N_{\mathbf{q}}+M^{\ast}_{\mathbf{q}}\right)\left(1+N_{\mathbf{p}-\mathbf{k}-\mathbf{q}}+M^{\ast}_{\mathbf{p}-\mathbf{k}-\mathbf{q}}\right)\nonumber\\ &\qquad +\:M_{\mathbf{k}}\left(N_{\mathbf{p}}+M^{\ast}_{\mathbf{p}}\right)\left(1+N_{\mathbf{q}}+M_{\mathbf{q}}\right)\left(1+N_{\mathbf{p}-\mathbf{k}-\mathbf{q}}+M_{\mathbf{p}-\mathbf{k}-\mathbf{q}}\right)\nonumber\\ &\qquad -\:M_{\mathbf{k}}\left(1+N_{\mathbf{p}}+M^{\ast}_{\mathbf{p}}\right)\left(N_{\mathbf{q}}+M_{\mathbf{q}}\right)\left(N_{\mathbf{p}-\mathbf{k}-\mathbf{q}}+M_{\mathbf{p}-\mathbf{k}-\mathbf{q}}\right)\;.\nonumber \\ 
\end{flalign}
\end{subequations}
%


\end{document}